\documentclass[english,notitlepage,pra,twocolumn]{revtex4-1}
\usepackage[T1]{fontenc}
\usepackage[latin9]{inputenc}
\usepackage{amsmath}
\usepackage{amssymb}
\usepackage{stackrel}
\usepackage{wasysym}
\usepackage{esint}
\usepackage{babel}
\begin{document}

\title{Universal relations for spin-orbit-coupled Fermi gases in two and
three dimensions}

\author{Cai-Xia Zhang}

\affiliation{Guangdong Provincial Key Laboratory of Quantum Engineering and Quantum
Materials, GPETR Center for Quantum Precision Measurement and SPTE,
South China Normal University, Guangzhou 510006, China}

\author{Shi-Guo Peng}
\email{pengshiguo@gmail.com}

\author{Kaijun Jiang}
\email{kjjiang@wipm.ac.cn}

\affiliation{State Key Laboratory of Magnetic Resonance and Atomic and Molecular
Physics, Wuhan Institute of Physics and Mathematics, Chinese Academy
of Sciences, Wuhan 430071, China}

\affiliation{Center for Cold Atom Physics, Chinese Academy of Sciences, Wuhan
430071, China}

\date{\today}
\begin{abstract}
We present a comprehensive derivation of a set of universal relations
for spin-orbit-coupled Fermi gases in three or two dimension, which
follow from the short-range behavior of the two-body physics. Besides
the adiabatic energy relations, the large-momentum distribution, the
grand canonical potential and pressure relation derived in our previous
work for three-dimensional systems {[}Phys. Rev. Lett. 120, 060408
(2018){]}, we further derive high-frequency tail of the radio-frequency
spectroscopy and the short-range behavior of the pair correlation
function. In addition, we also extend the derivation to two-dimensional
systems with Rashba type of spin-orbit coupling. To simply demonstrate
how the spin-orbit-coupling effect modifies the two-body short-range
behavior, we solve the two-body problem in the sub-Hilbert space of
zero center-of-mass momentum and zero total angular momentum, and
perturbatively take the spin-orbit-coupling effect into account at
short distance, since the strength of the spin-orbit coupling should
be much smaller than the corresponding scale of the finite range of
interatomic interactions. The universal asymptotic forms of the two-body
wave function at short distance are then derived, which do not depend
on the short-range details of interatomic potentials. We find that
new scattering parameters need to be introduced because of spin-orbit
coupling, besides the traditional $s$- and $p$-wave scattering length
(volume) and effective ranges. This is a general and unique feature
for spin-orbit-coupled systems. We show how these two-body parameters
characterize the universal relations in the presence of spin-orbit
coupling. This work probably shed light for understanding the profound
properties of the many-body quantum systems in the presence of the
spin-orbit coupling.
\end{abstract}
\maketitle

\section{Introduction}

Understanding strongly-interacting many-body systems is one of the
most daunting challenges in modern physics. Owing to the development
of the experimental technique, ultracold atomic gases acquire a high
degree of control and tunability in interatomic interaction, geometry,
purity, atomic species, and lattice constant (of optical lattices)
\cite{Bloch2008M,Giorgini2008T,Chin2010F,Kohler2006P,Gross2017Q}.
To date, ultracold quantum gases have emerged as a versatile platform
for exploring a broad variety of many-body phenomena as well as offering
numerous examples of interesting many-body states \cite{Qi2011T,Dalibard2011A,Abanin2018M}.
Unlike conventional electric gases in condensed matters, atomic quantum
gases are extremely dilute, and the mean distance between atoms is
usually very large (on the order of $\mu$m), while the range of interatomic
interactions is several orders smaller (on the order of several tens
of nm). Therefore, the two-body correlations characterize the key
properties of such many-body systems near scattering resonances, where
the two-body interactions are simply described by the scattering length
and become irrelevant to the specific form of interatomic potentials. 

A set of universal relations, following from the short-range behavior
of the two-body physics, govern some crucial features of ultracold
atomic gases, and provide powerful constraints on the behavior of
the system. Many of these relations were first derived by Shina Tan,
such as the adiabatic energy relation, energy theorem, general virial
theorem and pressure relation \cite{Tan2008E,Tan2008L,Tan2008G}.
Afterwards, more universal behaviors were obtained by others, such
as the radio-frequency (rf) spectroscopy, photoassociation, static
structure factors and so on \cite{Zwerger2011T}. All these relations
are characterized by the only universal quantity named\emph{ contact},
and therefore known as the contact theory. During past few years,
the concept of contact theory was further generalized to higher-partial-wave
interactions \cite{Yu2015U,Yoshida2015U,He2016C,Yoshida2016P,Peng2016L,Qin2016U,Zhang2017E,Zhang2017C}
as well as to low dimensions \cite{Barth2011T,Valiente2011U,Werner2012GF,Werner2012GB,Cui2016U,Cui2016H,Zhang2017S,Yin2018M,Peng2019U},
and more contacts appear when additional two-body parameters are involved.

The reason why the contact theory is significantly important in ultracold
atoms is attributed to its direct connection to the experimental measurements.
Some of the universal relations were experimentally confirmed, involving
various measurements of the contact itself. For two-component Fermi
gases with $s$-wave interactions, D. S. Jin's group measured the
contact according to three different methods, i.e., the momentum distribution,
photoemission spectroscopy, and rf spectroscopy, and tested the adiabatic
energy relation when the interatomic interaction was adiabatically
swept \cite{Stewart2010V}. The asymptotic behavior of the static
structure factor at large momentum was confirmed by C. J. Vale's group,
by using Bragg spectroscopy technique \cite{Kuhnle2010U,Hoinka2013P}.
Recently, the temperature evolution of the contact was resolved independently
by M. Zwierlein's group and C. J. Vale's group, especially the characteristic
behavior of the contact across the superfluid transition \cite{Mukherjee2019S,Carcy2019C}.
For single-component Fermi gases with $p$-wave interactions, the
feasibility of generalizing the contact theory for higher-partial-wave
scatterings was confirmed experimentally by Thywissen's group \cite{Luciuk2016E},
in which the anisotropic $p$-wave interaction was tuned according
to the magnetic vector \cite{Peng2014M}. Nowadays, the contact gradually
becomes one of fundamental concepts in ultracold atomic physics both
theoretically and experimentally.

In the past decade, the realizations of the spin-orbit (SO) coupling
in ultracold neutral atoms have sparked a great deal of interest \cite{Lin2011S,Cheuk2012S,Wang2012S,Wu2016R,Huang2016E,Chen2018S,Chen2018R,Zhang2019G}.
It provides an ideal platform on which to study novel quantum phenomena
resulted from SO coupling in a highly controllable and tunable way,
such as topological insulators and superconductors \cite{Qi2011T,Dalibard2011A},
and (spin) Hall effect \cite{Zhu2006S,Beeler2013T,Choi2013O}. Nevertheless,
it is still challenging to theoretically deal with the many-body correlations
for SO-coupled systems. Unlike the situation in condensed matters,
the intrinsic short-range feature of interatomic potentials is unchanged
for neutral atoms even in the presence of SO coupling. The natural
question may be raised, from the point of view of the contact theory,
as to whether the two-body physics could provide crucial constraints
on many-body behaviors of SO-coupled atomic systems. In addition,
it was pointed out that although the short-range feature remains,
the SO-coupling effect does modify the short-range behavior of the
two-body wave function \cite{Zhang2012M}. Therefore, the existence
and exact forms of universal relations for SO-coupled atomic systems
attract a great deal of attention. In \cite{Peng2018C}, we preliminarily
discussed some of the universal relations for three-dimensional (3D)
Fermi gases in the presence of 3D isotropic SO coupling. We proposed
a simple way to construct the short-range wave function, in which
the SO coupling effect could be taken into account perturbatively.
Since SO-coupling in general couples different partial waves of the
two-body scatterings, additional contact parameters appear in universal
relations. Before long, our theory was verified by different groups
near $s$-wave resonances \cite{Zhang2018U,Jie2018U}. 

So far, the generalization of the contact theory in the presence of
SO-coupling is mostly discussed in 3D, while the derivation of these
universal relations is still elusive in two-dimensional (2D) systems.
The short-range behavior of the two-body physics in 2D is different
from that in 3D: the two-body wave function in 3D is power-law divergent,
while one has to deal with the logarithmic divergence in 2D. From
the point of view of the contact theory, different short-range correlations
in two-body physics result in different forms of universal relations.
Therefore, it requires a direct extension to 2D in the similar manner
as in 3D in the presence of SO coupling. 

The purpose of this article is to present a comprehensive derivation
of universal relations for SO-coupled Fermi gases. Besides the adiabatic
energy relations, the large-momentum distribution, the grand canonical
potential and pressure relation derived in our previous work for 3D
systems \cite{Peng2018C}, we further derive high-frequency tail of
the rf spectroscopy and the short-range behavior of the pair correlation
function. Then we generalize the derivation of universal relations
for 3D systems to 2D case with Rashba SO coupling in a similar way.
For the convenience of the presentation, we still construct the short-range
behavior of the two-body wave function in the sub-Hilbert space of
zero center-of-mass (c.m.) momentum and zero total angular momentum
as before, and then only $s$- and $p$-wave scatterings are coupled
\cite{Peng2018C,Cui2012M,Wu2013S}. Our results show that the SO coupling
introduces a new contact and modifies the universal relations of many-body
systems. 

The remainder of this paper is organized as follows. In the next section,
we present the derivations of the short-range behavior of two-body
wave functions for SO-coupled Fermi gases in three and two dimensions,
respectively. Subsequently, with the short-range behavior of the two-body
wave functions in hands, we derive a set of universal relations for
a 3D SO-coupled Fermi gases in Sec. III, and then generalize them
to 2D SO-coupled Fermi gases in Sec. IV, including adiabatic energy
relations, asymptotic behavior of the large-momentum distribution,
the high-frequency behavior of the rf response, short-range behavior
of the pair correlation function, grand canonical potential and pressure
relation. Finally, the main results are summarized in Sec. V. 

\section{Universal short-range behavior of two-body wave functions}

The ultracold atomic gases are dilute, while the range of interatomic
potentials is extremely small. When two fermions get close enough
to interact with each other, they usually far away from the others.
If only these two-body correlations are taken into account, some key
properties of many-body systems are characterized by the short-range
two-body physics, which is the basic idea of the contact theory. In
this section, we are going to discuss the short-range behavior of
two-body wave functions for 3D Fermi gases in the presence of 3D SO
coupling and 2D Fermi gases in the presence of 2D SO coupling, respectively.
Let us consider spin-half SO-coupled Fermi gases, and the Hamiltonian
of a single fermion is modeled as

\begin{equation}
\hat{\mathcal{H}}_{1}=\frac{\hbar^{2}\mathbf{\hat{k}_{1}^{\mathrm{2}}}}{2M}+\frac{\hbar^{2}\lambda}{M}\hat{\chi}+\frac{\hbar^{2}\lambda^{2}}{2M},\label{eq:s1}
\end{equation}
where $\mathbf{\hat{k}}_{1}=-i\boldsymbol{\nabla}$ is the single-particle
momentum operator, $M$ is the atomic mass, $\hbar$ is the Planck's
constant divided by $2\pi$. Here, the SO coupling is described by
the term $\hbar^{2}\lambda\hat{\chi}/M$ with the strength $\lambda>0$,
and $\hat{\chi}$ takes the isotropic form of $\hat{{\bf k}}_{1}\cdot\hat{\boldsymbol{\sigma}}$
in 3D or the Rashba form of $\hat{\boldsymbol{\sigma}}\times\mathbf{\hat{k}}_{1}\cdot\mathbf{\hat{n}}$
in 2D \cite{Bychkov1984O}, where $\hat{\boldsymbol{\sigma}}$ is
the Pauli operator, and $\mathbf{\hat{n}}$ is the unit vector perpendicular
to the ($x-y$) plane. 

Because of SO coupling, the orbital angular momentum of the relative
motion of two fermions is no longer conserved, and then all the partial-wave
scatterings are coupled \cite{Cui2012M}. Fortunately, the c.m. momentum
${\bf K}$ of two fermions is still conserved as well as the total
angular momentum ${\bf J}$. For simplicity, we may reasonably focus
on the two-body problem in the subspace of $\mathbf{K}=0$ and $\mathbf{J}=0$,
and then only $s$- and $p$-wave scatterings are involved \cite{Cui2012M,Wu2013S}.
Consequently, the Hamiltonian of two spin-half fermions can be written
as

\begin{equation}
\mathcal{\hat{H}}_{2}=\frac{\hbar^{2}\mathbf{\hat{k}}^{2}}{M}+\frac{\hbar^{2}\lambda}{M}\hat{Q}\left(\mathbf{r}\right)+\frac{\hbar^{2}\lambda^{2}}{M}+V\left({\bf r}\right),\label{eq:t2}
\end{equation}
where $\mathbf{\hat{k}}=\left(\mathbf{\hat{k}}_{2}-\mathbf{\hat{k}}_{1}\right)/2$
is the momentum operator for the relative motion $\mathbf{r=r_{2}-r_{1}}$,
$V\left({\bf r}\right)$ is the short-range interatomic interaction
with a finite range $\epsilon$, $\hat{Q}\left(\mathbf{r}\right)=\left(\mathbf{\hat{\boldsymbol{\sigma}}_{2}-\hat{\boldsymbol{\sigma}}_{1}}\right)\cdot\mathbf{\hat{k}}$
in 3D or $\hat{Q}\left(\mathbf{r}\right)=\left(\mathbf{\hat{\boldsymbol{\sigma}}_{2}-\hat{\boldsymbol{\sigma}}_{1}}\right)\times\mathbf{\hat{k}}\cdot\mathbf{\hat{n}}$
in 2D, and $\hat{\boldsymbol{\sigma}}_{i}$ is the spin operator of
the $i$th atom. In the follows, let us consider the two-body problems
in the 3D systems with 3D SO coupling and 2D systems with 2D SO coupling,
respectively.

\subsection{For 3D systems with 3D SO coupling}

In the subspace of $\mathbf{K}=0$ and $\mathbf{J}=0$, we may choose
the common eigenstates of the total Hamiltonian $\mathcal{\hat{H}}_{2}$
and total angular momentum $\mathbf{J}\text{(\ensuremath{=0})}$ as
the basis of Hilbert space, which take the forms of

\begin{eqnarray}
\Omega_{0}\left(\mathbf{\hat{r}}\right) & = & Y_{00}\left(\mathbf{\hat{r}}\right)\left|S\right\rangle ,\label{eq:bas3}\\
\Omega_{1}\left(\mathbf{\hat{r}}\right) & = & -\frac{i}{\sqrt{3}}\left[Y_{1-1}\left(\mathbf{\hat{r}}\right)\left|\uparrow\uparrow\right\rangle \right.\nonumber \\
 &  & \left.+Y_{11}\left(\mathbf{\hat{r}}\right)\left|\downarrow\downarrow\right\rangle -Y_{10}\left(\mathbf{\hat{r}}\right)\left|T\right\rangle \right],\label{eq:bas31}
\end{eqnarray}
where $Y_{lm}\left(\mathbf{\hat{r}}\right)$ is the spherical harmonics,
$\mathbf{\hat{r}}\equiv\left(\theta,\varphi\right)$ denotes the angular
degree of freedom of the coordinate $\mathbf{r}$, and $\left|S\right\rangle =\left(\left|\uparrow\downarrow\right\rangle -\left|\downarrow\uparrow\right\rangle \right)/\sqrt{2}$
and $\left\{ \left|\uparrow\uparrow\right\rangle ,\left|\downarrow\downarrow\right\rangle ,\left|T\right\rangle =\left(\left|\uparrow\downarrow\right\rangle +\left|\downarrow\uparrow\right\rangle \right)/\sqrt{2}\right\} $
are the spin-singlet and spin-triplet states with total spin $S=0$
and $1$, respectively. Then the two-body wave function can formally
be written in the basis of $\left\{ \Omega_{0}\left(\mathbf{\hat{r}}\right),\Omega_{1}\left(\mathbf{\hat{r}}\right)\right\} $
as

\begin{equation}
\Psi\left(\mathbf{r}\right)=\psi_{0}\left(r\right)\Omega_{0}\left(\mathbf{\hat{r}}\right)+\psi_{1}\left(r\right)\Omega_{1}\left(\mathbf{\hat{r}}\right),\label{eq:2bw}
\end{equation}
where $\psi_{i}\left(r\right)\,\left(i=0,1\right)$ is the radial
part of the wave function. Note that we here consider an isotropic
$p$-wave interaction and the radial wave function $\psi_{1}\left(r\right)$
is identical for three scattering channels, i.e., $m=0,\pm1$. 

Typically, the SO coupling strength (of the order $\mu$m$^{-1}$)
is pretty small compared to the inverse of the interaction range (of
the order nm$^{-1}$) \cite{Cheuk2012S,Wang2012S}, i.e., $\lambda\ll\epsilon^{-1}$.
Moreover, in the low-energy scattering limit, the relative momentum
$k=\sqrt{ME/\hbar^{2}}$ is also much smaller than $\epsilon^{-1}$.
Thus, when two fermions get as close as the range of the interaction,
i.e., $r\sim\epsilon$, the SO coupling can be treated as perturbation
as well as the energy. We assume that the two-body wave function may
take the form of the following ansatz \cite{Peng2018C}

\begin{equation}
\Psi\left(\mathbf{r}\right)\approx\phi\left(\mathbf{r}\right)+k^{2}f\left(\mathbf{r}\right)-\lambda g\left(\mathbf{r}\right),\label{eq:ansatz}
\end{equation}
as the distance of two fermions approaches $\epsilon$. Here, we keep
up to the first-order terms of the energy ($k^{2}$) and SO coupling
strength ($\lambda$). The advantage of this ansatz is that the functions
$\phi\left(\mathbf{r}\right)$, $f\left(\mathbf{r}\right)$, and $g\left(\mathbf{r}\right)$
are all independent on $k^{2}$ and $\lambda$. These functions are
determined only by the short-range details of the interaction, and
thus characterize the intrinsic properties of the interatomic potential.
We expect that in the absence of SO coupling the conventional scattering
length or volume is included in the zero-order term $\phi\left(\mathbf{r}\right)$,
while the effective range is included in $f\left(\mathbf{r}\right)$,
the coefficient of the first-order term of $k^{2}$. Interestingly,
we may anticipate that new scattering parameters resulted from SO
coupling appear in the first-order term of $\lambda$ {[}in $g\left(r\right)${]}.
Conveniently, more scattering parameters may be introduced if higher-order
terms of $k^{2}$ and $\lambda$ are perturbatively considered.

Inserting the ansatz (\ref{eq:ansatz}) into the Schr\"{o}dinger
equation $\mathcal{\hat{H}}_{2}\Psi\left(\mathbf{r}\right)=E\Psi\left(\mathbf{r}\right)$,
and comparing the corresponding coefficients of $k^{2}$ and $\lambda$
, we find

\begin{eqnarray}
\left[-\nabla^{2}+\frac{MV\left({\bf r}\right)}{\hbar^{2}}\right]\phi\left(\mathbf{r}\right) & = & 0,\label{eq:f1}\\
\left[-\nabla^{2}+\frac{MV\left({\bf r}\right)}{\hbar^{2}}\right]f\left(\mathbf{r}\right) & = & \phi\left(\mathbf{r}\right),\label{eq:f2}\\
\left[-\nabla^{2}+\frac{MV\left({\bf r}\right)}{\hbar^{2}}\right]g\left(\mathbf{r}\right) & = & \hat{Q}\left(\mathbf{r}\right)\phi\left(\mathbf{r}\right).\label{eq:f3}
\end{eqnarray}
These coupled equations can easily be solved for $r>\epsilon$, and
we obtain
\begin{eqnarray}
\phi\left({\bf r}\right) & = & \alpha_{0}\left(\frac{1}{r}-\frac{1}{a_{0}}\right)\Omega_{0}\left(\hat{{\bf r}}\right)\nonumber \\
 &  & +\alpha_{1}\left(\frac{1}{r^{2}}-\frac{1}{3a_{1}}r\right)\Omega_{1}\left(\hat{{\bf r}}\right)+O\left(r^{2}\right),\label{eq:f4}\\
f\left({\bf r}\right) & = & \alpha_{0}\left(\frac{1}{2}b_{0}-\frac{1}{2}r\right)\Omega_{0}\left(\hat{{\bf r}}\right)\nonumber \\
 &  & +\alpha_{1}\left(\frac{1}{2}+\frac{b_{1}}{6}r\right)\Omega_{1}\left(\hat{{\bf r}}\right)+O\left(r^{2}\right),\label{eq:f5}\\
g\left({\bf r}\right) & = & -\alpha_{1}u\Omega_{0}\left(\hat{{\bf r}}\right)-\alpha_{0}\left(1+vr\right)\Omega_{1}\left(\hat{{\bf r}}\right)+O\left(r^{2}\right),\label{eq:f6}
\end{eqnarray}
where $\alpha_{0}$ and $\alpha_{1}$ are two complex superposition
coefficients, $a_{i}$ and $b_{i}$ are $s$-wave scattering length
and effective range for $i=0$, and $p$-wave scattering volume and
effective range for $i=1$, respectively. Interestingly, two new scattering
parameters $u$ and $v$ are involved as we anticipate. They are corrections
from SO coupling to the short-range behavior of the two-body wave
function in $s$- and $p$-wave channels, respectively. 

In the absence of SO coupling, if atoms are initially prepared near
an $s$-wave resonance, the contribution from the $p$-wave channel
could be ignored, and we have $\alpha_{1}\approx0$. Naturally, the
two-body wave function $\Psi\left({\bf r}\right)$ reduces to the
known $s$-wave form of (up to a constant $\alpha_{0}$)
\begin{equation}
\Psi\left({\bf r}\right)=\left(\frac{1}{r}-\frac{1}{a_{0}}+\frac{b_{0}k^{2}}{2}-\frac{k^{2}}{2}r\right)\Omega_{0}\left(\hat{{\bf r}}\right)+O\left(r^{2}\right)\label{eq:f7}
\end{equation}
at short distance $r\apprge\epsilon$. Subsequently, when SO coupling
is switched on near the $s$-wave resonance, a considerable $p$-wave
contribution is involved, and the two-body wave function becomes
\begin{multline}
\Psi_{s}\left({\bf r}\right)=\left(\frac{1}{r}-\frac{1}{a_{0}}+\frac{b_{0}k^{2}}{2}-\frac{k^{2}}{2}r\right)\Omega_{0}\left(\hat{{\bf r}}\right)\\
+\left(1+vr\right)\lambda\Omega_{1}\left(\hat{{\bf r}}\right)+O\left(r^{2}\right),\label{eq:f8}
\end{multline}
which recovers the modified Bethe-Peierls boundary condition of \cite{Zhang2012M}
by noticing $\Omega_{0}\left(\hat{{\bf r}}\right)=\left|S\right\rangle /\sqrt{4\pi}$
and $\Omega_{1}\left(\hat{{\bf r}}\right)=-i\left(\hat{\boldsymbol{\sigma}}_{2}-\hat{\boldsymbol{\sigma}}_{1}\right)\cdot\left({\bf r}/r\right)\left|S\right\rangle /\sqrt{16\pi}$.
We can see that the parameter $v$ characterizes the hybridization
of the $p$-wave component into the $s$-wave scattering due to SO
coupling. If atoms are initially prepared near a $p$-wave resonance
without SO coupling, the $s$-wave scattering could be ignored, then
we have $\alpha_{0}\approx0$. The two-body wave function $\Psi\left({\bf r}\right)$
takes the known $p$-wave form at short distance, i.e., 
\begin{equation}
\Psi\left({\bf r}\right)=\left(\frac{1}{r^{2}}-\frac{1}{3a_{1}}r+\frac{k^{2}}{2}+\frac{b_{1}k^{2}}{6}r\right)\Omega_{1}\left(\hat{{\bf r}}\right)+O\left({\bf r}^{2}\right).\label{eq:f9}
\end{equation}
 In the presence of SO coupling near the $p$-wave resonance, an $s$-wave
component is introduced, and the two-body wave function becomes
\begin{multline}
\Psi_{p}\left({\bf r}\right)=\left[\frac{1}{r^{2}}+\frac{k^{2}}{2}+\left(-\frac{1}{3a_{1}}+\frac{b_{1}k^{2}}{6}\right)r\right]\Omega_{1}\left(\hat{{\bf r}}\right)\\
+u\lambda\Omega_{0}\left(\hat{{\bf r}}\right)+O\left({\bf r}^{2}\right)\label{eq:f10}
\end{multline}
at short distance. We can see that the parameter $u$ describes the
hybridization of the $s$-wave component into the $p$-wave scattering
due to SO coupling. In general, both $s$- and $p$-wave scatterings
exist between atoms in the absence of SO coupling. Therefore, when
SO coupling is introduced, the two-body wave function is generally
the arbitrary superposition of Eqs.(\ref{eq:f8}) and (\ref{eq:f10}),
and can be written as\begin{widetext}
\begin{multline}
\Psi_{3D}\left({\bf r}\right)=\alpha_{0}\left(\frac{1}{r}-\frac{1}{a_{0}}+\frac{b_{0}k^{2}}{2}+\frac{\alpha_{1}}{\alpha_{0}}u\lambda-\frac{k^{2}}{2}r\right)\Omega_{0}\left(\hat{{\bf r}}\right)\\
+\alpha_{1}\left[\frac{1}{r^{2}}+\frac{k^{2}}{2}+\frac{\alpha_{0}}{\alpha_{1}}\lambda+\left(-\frac{1}{3a_{1}}+\frac{b_{1}k^{2}}{6}+\frac{\alpha_{0}}{\alpha_{1}}v\lambda\right)r\right]\Omega_{1}\left(\hat{{\bf r}}\right)+O\left(r^{2}\right)\label{eq:f11}
\end{multline}
\end{widetext} at short distance $r\apprge\epsilon$. Eq.(\ref{eq:f11})
can be treated as the short-range boundary condition for two-body
wave functions in 3D in the presence of 3D SO coupling, when both
$s$- and $p$-wave interactions are considered.

\subsection{For 2D systems with 2D SO coupling}

Let us consider two spin-half fermions scattering in the $x-y$ plane.
We easily find that the total angular momentum ${\bf J}$ perpendicular
to the $x-y$ plane is conserved as well as the c.m. momentum ${\bf K}$.
Therefore, we may still focus on the two-body problem in the subspace
of $\mathbf{K=0}$ and $\mathbf{J=0}$, which is spanned by the following
three orthogonal basis

\begin{eqnarray}
\Omega_{0}\left(\varphi\right) & = & \frac{1}{\sqrt{2\pi}}\left|S\right\rangle ,\label{eq:bas2}\\
\Omega_{-1}\left(\varphi\right) & = & \frac{e^{-i\varphi}}{\sqrt{2\pi}}\left|\uparrow\uparrow\right\rangle ,\label{eq:bas21}\\
\Omega_{1}\left(\varphi\right) & = & \frac{e^{i\varphi}}{\sqrt{2\pi}}\left|\downarrow\downarrow\right\rangle ,\label{eq:bas22}
\end{eqnarray}
where $\varphi$ is the azimuthal angle of the relative coordinate
${\bf r}$. Then the two-body wave function can formally be expanded
as
\begin{equation}
\Psi\left({\bf r}\right)=\sum_{m=0,\pm1}\psi_{m}\left(r\right)\Omega_{m}\left(\varphi\right),\label{eq:2Dwf}
\end{equation}
and $\psi_{m}\left(r\right)$ is the radial wave function. Analogously,
the strength of SO coupling as well as the energy can be taken into
account perturbatively at short distance. We assume that the two-body
wave function has the form of the ansatz (\ref{eq:ansatz}), and the
corresponding functions to be determined can easily be solved out
from the Schr\"{o}dinger equation outside the range of the interatomic
potential, i.e., $r\apprge\epsilon$. After straightforward algebra,
we obtain
\begin{multline}
\phi\left({\bf r}\right)=\alpha_{0}\left(\ln\frac{r}{2a_{0}}+\gamma\right)\Omega_{0}\left(\varphi\right)\\
+\left(\frac{1}{r}-\frac{\pi}{4a_{1}}r\right)\sum_{m=\pm1}\alpha_{m}\Omega_{m}\left(\varphi\right)+O\left(r^{2}\right),
\end{multline}
\begin{multline}
f\left({\bf r}\right)=-\alpha_{0}\left(\frac{\pi}{4}b_{0}+\frac{1}{4}r^{2}\ln\frac{r}{2a_{0}}\right)\Omega_{0}\left(\varphi\right)\\
+\left(\frac{1-2\gamma}{4}r-\frac{1}{2}r\ln\frac{r}{2b_{1}}\right)\sum_{m=\pm1}\alpha_{m}\Omega_{m}\left(\varphi\right)+O\left(r^{2}\right),
\end{multline}
\begin{multline}
g\left({\bf r}\right)=-\left(\sum_{m=\pm1}\alpha_{m}\right)u\Omega_{0}\left(\varphi\right)\\
-\alpha_{0}\left(vr+\frac{r}{\sqrt{2}}\ln\frac{r}{2b_{1}}\right)\sum_{m=\pm1}\Omega_{m}\left(\varphi\right)+O\left(r^{2}\right)
\end{multline}
 for $r\apprge\epsilon$, where $\gamma$ is Euler's constant, $\alpha_{m}$
$\left(m=0,\pm1\right)$ is complex superposition coefficients, $a_{m}$
and $b_{m}$ are $s$-wave scattering length and effective range for
$m=0$, and $p$-wave scattering area and effective range for $\left|m\right|=1$,
respectively. Here, we have assumed that the $p$-wave interaction
is isotropic and thus is the same in $m=\pm1$ channels, and applied
the $p$-wave effective-range expansion of the scattering phase shift,
i.e., $k^{2}\cot\delta_{1}=-1/a_{1}+2k^{2}\ln\left(kb_{1}\right)/\pi$
\cite{Zhang2018H}. We find that two new scattering parameters are
similarly introduced, and they demonstrate the hybridization of $s$-
and $p$-wave scattering in the presence of Rashba SOC in 2D. Finally,
the asymptotic form of the two-body wave function at short distance
can be written as\begin{widetext}
\begin{multline}
\Psi_{2D}\left({\bf r}\right)=\alpha_{0}\left[\ln\frac{r}{2a_{0}}+\gamma-\frac{\pi}{4}b_{0}k^{2}+\left(\sum_{m=\pm1}\frac{\alpha_{m}}{\alpha_{0}}\right)u\lambda-\frac{k^{2}}{4}r^{2}\ln\frac{r}{2a_{0}}\right]\Omega_{0}\left(\varphi\right)\\
+\sum_{m=\pm1}\alpha_{m}\left[\frac{1}{r}+\left(-\frac{\pi}{4a_{1}}+\frac{1-2\gamma}{4}k^{2}+\frac{\alpha_{0}}{\alpha_{m}}v\lambda\right)r+\left(-\frac{k^{2}}{2}+\frac{\alpha_{0}}{\alpha_{m}}\frac{\lambda}{\sqrt{2}}\right)r\ln\frac{r}{2b_{1}}\right]\Omega_{m}\left(\varphi\right)+O\left(r^{2}\right)\label{eq:u2d}
\end{multline}
\end{widetext}for $r\apprge\epsilon$. It is apparent that $\Psi_{2D}\left({\bf r}\right)$
naturally decouples to the $s$- and $p$-wave short-range boundary
conditions in the absence of SO coupling. However, Rashba SO coupling
mixes the $s$- and $p$-wave scatterings, and two new scattering
parameters $u$ and $v$ are introduced. We should note that the short-range
behaviors of the two-body wave function, i.e., Eqs.(\ref{eq:f11})
and (\ref{eq:u2d}), are universal and does not depend on the specific
form of interatomic potentials. 

\section{Universal relations in the presence of isotropic 3D SO coupling}

In the previous section, we have discussed the two-body problem in
the presence of SO coupling, and obtained the short-range behaviors
of the two-body wave functions. Then, we are ready to consider Tan's
universal relations of SO-coupled many-body systems, if only two-body
correlations are taken into account. Owing to the short-range property
of interactions between neutral atoms, when two fermions ($i$ and
$j$) get as close as the range of interatomic potentials, all the
other atoms are usually far away. In this case, the many-body wave
functions approximately take the forms of Eq.(\ref{eq:f11}) in 3D
systems with 3D SO coupling, when the fermions $i$ and $j$ approach
to each other. We need to pay attention that the arbitrary superposition
coefficient $\alpha_{m}\left({\bf X}\right)$ then becomes the functions
of the c.m. coordinates of the pair $\left(i,j\right)$ as well as
those of the rest of the fermions, which we include into the variable
${\bf X}$. In the follows, we derive a set of universal relations
for SO-coupled many-body systems by using Eqs.(\ref{eq:f11}) for
3D SO-coupled Fermi gases. These relations include adiabatic energy
relations, the large-momentum behavior of the momentum distribution,
the high-frequency tail of the rf spectroscopy, the short-range behavior
of the pair correlation function, the grand canonical potential and
pressure relation. Let us consider a strongly interacting two-component
Fermi gases with total atom number $N$. For simplicity, we consider
the case with $b_{0}\approx0$ for broad $s$-wave resonances in the
follows.

\subsection{Adiabatic energy relations}

In order to investigate how the energy varies with the two-body interaction,
let us consider two many-body wave functions $\Psi$ and $\Psi^{\prime}$,
corresponding to different interatomic interaction strengths. They
satisfy the Schr\"{o}dinger equation with different energies

\begin{eqnarray}
\sum_{i=1}^{N}\hat{\mathcal{H}}_{1}^{(i)}\Psi & = & E\Psi,\label{eq:tas2}\\
\sum_{i=1}^{N}\hat{\mathcal{H}}_{1}^{(i)}\Psi^{\prime} & = & E^{\prime}\Psi^{\prime},\label{eq:tas3}
\end{eqnarray}
if there is not any pair of atoms within the range of the interaction,
where $\hat{\mathcal{H}}_{1}^{(i)}$ denotes the single-atom Hamiltonian
(\ref{eq:s1}) for the $i$th fermion. By subtracting $[\ref{eq:tas3}]^{*}\times\Psi$
from $\Psi^{\prime*}\times\left[\ref{eq:tas2}\right]$, and integrating
over the domain $\mathcal{D_{\epsilon}}$, the set of all configurations
$\left({\bf r}_{i},{\bf r}_{j}\right)$ in which $r=\left|{\bf r}_{i}-{\bf r}_{j}\right|>\epsilon$,
we arrive at

\begin{multline}
\left(E-E^{\prime}\right)\int_{\mathcal{D_{\epsilon}}}\prod_{i=1}^{N}d\mathbf{r}_{i}\Psi^{\prime*}\Psi=\\
-\frac{\hbar^{2}}{M}\mathcal{N}\int_{r>\epsilon}d{\bf X}d\mathbf{r}\left[\Psi^{\prime*}\nabla_{\mathbf{r}}^{2}\Psi-\left(\nabla_{\mathbf{r}}^{2}\Psi^{\prime*}\right)\Psi\right]\\
+\frac{\hbar^{2}\lambda}{M}\mathcal{N}\int_{r>\epsilon}d{\bf X}d\mathbf{r}\left[\Psi^{\prime*}\left(\hat{Q}\Psi\right)-\left(\hat{Q}\Psi^{\prime}\right)^{*}\Psi\right],\label{eq:tas4}
\end{multline}
where $\mathcal{N}=N\left(N-1\right)/2$ is the number of all the
possible ways to pair atom. Using the Gauss\textquoteright{} theorem,
the first term on the right-hand side (RHS) can be written as 

\begin{align}
 & -\frac{\hbar^{2}}{M}\mathcal{N}\int_{r>\epsilon}d{\bf X}d\mathbf{r}\left[\Psi^{\prime*}\nabla_{\mathbf{r}}^{2}\Psi-\left(\nabla_{\mathbf{r}}^{2}\Psi^{\prime*}\right)\Psi\right]\nonumber \\
= & -\frac{\hbar^{2}}{M}\mathcal{N}\varoiint_{r=\epsilon}\left[\Psi^{\prime*}\nabla_{\mathbf{r}}\Psi-\left(\nabla_{\mathbf{r}}\Psi^{\prime*}\right)\Psi\right]\cdot\hat{\mathbf{n}}d\mathcal{S}\nonumber \\
= & \frac{\hbar^{2}\epsilon^{2}}{M}\mathcal{N}\stackrel[i=0]{1}{\sum}\int d{\bf X}\left(\psi_{i}^{\prime*}\frac{\partial}{\partial r}\psi_{i}-\psi_{i}\frac{\partial}{\partial r}\psi_{i}^{\prime*}\right)_{r=\epsilon},\label{eq:tas5}
\end{align}
where $\mathcal{S}$ is the surface in which the distance between
the two atoms in the pair ($i,j$) is $\epsilon$ with, $\hat{\mathbf{n}}$
is the direction normal to $\mathcal{S}$ but opposite to the radial
direction, and $\psi_{0}$ ($\psi_{1}$) is the $s$-wave ($p$-wave)
component of the radial two-body wave function. In addition, for the
second term on the RHS of Eq.(\ref{eq:tas4}), we have

\begin{equation}
\hat{Q}\left(\mathbf{r}\right)\Psi=-\frac{2}{r^{2}}\frac{\partial}{\partial r}\left(r^{2}\psi_{1}\right)\Omega_{0}\left(\mathbf{\hat{r}}\right)+2\frac{\partial\psi_{0}}{\partial r}\Omega_{1}\left(\mathbf{\hat{r}}\right),
\end{equation}
then it becomes 

\begin{multline}
\frac{\hbar^{2}\lambda}{M}\mathcal{N}\int_{r>\epsilon}d\mathbf{X}d\mathbf{r}\left[\Psi^{\prime*}\left(\hat{Q}\left(\mathbf{r}\right)\Psi\right)-\left(\hat{Q}\left(\mathbf{r}\right)\Psi^{\prime}\right)^{*}\Psi\right]\\
=\frac{2\lambda\hbar^{2}\epsilon^{2}}{M}\mathcal{N}\int d\mathbf{X}\left(\psi_{0}^{\prime*}\psi_{1}-\psi_{1}^{\prime*}\psi_{0}\right)_{r=\epsilon}.\label{eq:tas5-1}
\end{multline}
Combining Eqs.(\ref{eq:tas4}), (\ref{eq:tas5}) and (\ref{eq:tas5-1}),
we obtain
\begin{multline}
\left(E-E^{\prime}\right)\int_{\mathcal{D_{\epsilon}}}\prod_{i=1}^{N}d\mathbf{r}_{i}\Psi^{\prime*}\Psi\\
=\frac{\hbar^{2}\epsilon^{2}}{M}\mathcal{N}\stackrel[i=0]{1}{\sum}\int d{\bf X}\left(\psi_{i}^{\prime*}\frac{\partial}{\partial r}\psi_{i}-\psi_{i}\frac{\partial}{\partial r}\psi_{i}^{\prime*}\right)_{r=\epsilon}\\
+\frac{2\lambda\hbar^{2}\epsilon^{2}}{M}\mathcal{N}\int d\mathbf{X}\left(\psi_{0}^{\prime*}\psi_{1}-\psi_{1}^{\prime*}\psi_{0}\right)_{r=\epsilon}.\label{eq:tas5-3}
\end{multline}
Inserting the asymptotic form of the many-body wave function Eq.(\ref{eq:f11})
into Eq.(\ref{eq:tas5-3}), and letting $E^{\prime}\rightarrow E$
and $\Psi^{\prime}\rightarrow\Psi$, we find

\begin{multline}
\delta E\cdot\int_{\mathcal{D_{\epsilon}}}\stackrel[i=1]{N}{\prod}d\mathbf{r}_{i}\left|\Psi\right|^{2}=-\frac{\hbar^{2}}{M}\left(\mathcal{I}_{a}^{\left(0\right)}-\lambda\mathcal{I_{\lambda}}\right)\delta a_{0}^{-1}\\
-\frac{\hbar^{2}\mathcal{I}_{a}^{\left(1\right)}}{M}\delta a_{1}^{-1}+\frac{\mathcal{E}_{1}}{2}\delta b_{1}+\frac{3\lambda\hbar^{2}}{2M}\mathcal{I_{\lambda}}\delta v\\
-\frac{\lambda\hbar^{2}}{M}\left(2\lambda\mathcal{I}_{a}^{\left(1\right)}-\frac{1}{2}\mathcal{I_{\lambda}}\right)\delta u+\left(\frac{1}{\epsilon}+\frac{b_{1}}{2}\right)\mathcal{I}_{a}^{\left(1\right)}\delta E,\label{eq:tas7}
\end{multline}
where

\begin{eqnarray}
\mathcal{I}_{a}^{\left(m\right)} & = & \mathcal{N}\int d\mathbf{X}\left|\alpha_{m}\left(\mathbf{X}\right)\right|^{2},\label{eq:tas12}\\
\mathcal{E}_{m} & = & \mathcal{N}\int d\mathbf{X}\alpha_{m}^{*}\left(\mathbf{X}\right)\left[E-\hat{T}(\mathbf{X})\right]\alpha_{m}\left(\mathbf{X}\right),\label{eq:tas13}\\
\mathcal{I_{\lambda}} & = & \mathcal{N}\int d\mathbf{X}\alpha_{0}^{*}\left(\mathbf{X}\right)\alpha_{1}\left(\mathbf{X}\right)+\mathrm{c.c.},\label{eq:tas12-2}\\
\mathcal{E}_{\lambda} & = & \mathcal{N}\int d\mathbf{X}\alpha_{0}^{*}\left(\mathbf{X}\right)\left[E-\hat{T}(\mathbf{X})\right]\alpha_{1}\left(\mathbf{X}\right)+\mathrm{c.c.}\label{eq:tas13-1}
\end{eqnarray}
for $m=0,1$, and $\hat{T}(\mathbf{X})$ is the kinetic operator including
the c.m. motion of the pair $\left(i,j\right)$ and those of all the
rest fermions. Using the normalization of the many-body wave function
(see appendix A)

\begin{multline}
\int_{\mathcal{D_{\epsilon}}}\stackrel[i=1]{N}{\prod}d\mathbf{r}_{i}\left|\Psi\right|^{2}=1+\left(\frac{1}{\epsilon}+\frac{b_{1}}{2}\right)\mathcal{I}_{a}^{\left(1\right)},\label{eq:tas7-1}
\end{multline}
we can further simplify Eq. (\ref{eq:tas7}) as

\begin{multline}
\delta E=-\frac{\hbar^{2}}{M}\left(\mathcal{I}_{a}^{\left(0\right)}-\lambda\mathcal{I_{\lambda}}\right)\delta a_{0}^{-1}-\frac{\hbar^{2}\mathcal{I}_{a}^{\left(1\right)}}{M}\delta a_{1}^{-1}\\
+\frac{\mathcal{E}_{1}}{2}\delta b_{1}+\frac{3\lambda\hbar^{2}\mathcal{I_{\lambda}}}{2M}\delta v+\frac{\lambda\hbar^{2}}{2M}\left(\mathcal{I_{\lambda}}-4\lambda\mathcal{I}_{a}^{\left(1\right)}\right)\delta u,
\end{multline}
which yields the following set of adiabatic energy relations

\begin{eqnarray}
\frac{\partial E}{\partial a_{0}^{-1}} & = & -\frac{\hbar^{2}}{M}\left(\mathcal{I}_{a}^{\left(0\right)}-\lambda\mathcal{I_{\lambda}}\right),\label{eq:taz9}\\
\frac{\partial E}{\partial a_{1}^{-1}} & = & -\frac{\hbar^{2}\mathcal{I}_{a}^{\left(1\right)}}{M},\label{eq:taz11}\\
\frac{\partial E}{\partial b_{1}} & = & \frac{\mathcal{E}_{1}}{2},\label{eq:taz12}\\
\frac{\partial E}{\partial u} & = & \frac{\lambda\hbar^{2}}{2M}\left(\mathcal{I_{\lambda}}-4\lambda\mathcal{I}_{a}^{\left(1\right)}\right),\label{eq:taz15}\\
\frac{\partial E}{\partial v} & = & \frac{3\lambda\hbar^{2}\mathcal{I_{\lambda}}}{2M}.\label{eq:taz16}
\end{eqnarray}
Interestingly, two additional new adiabatic energy relations appear,
i.e. Eqs. (\ref{eq:taz15}) and (\ref{eq:taz16}), which originate
from new scattering parameters introduced by SO coupling. These relations
demonstrate how the macroscopic internal energy of an SO-coupled many-body
system varies with microscopic two-body scattering parameters. 

\subsection{Tail of the momentum distribution at large $q$}

Let us then study the asymptotic behavior of the large momentum distribution
for a many-body system with $N$ fermions. The momentum distribution
of the $i$th fermion is defined as

\begin{equation}
n_{i}\left(\mathbf{q}\right)=\int\underset{t\neq i}{\prod}d\mathbf{r}_{t}\left|\tilde{\Psi}_{i}\left(\mathbf{q}\right)\right|^{2},\label{eq:lm1}
\end{equation}
where $\tilde{\Psi}_{i}\left(\mathbf{q}\right)\equiv\int d\mathbf{r}_{i}\Psi_{3D}e^{-i\mathbf{q}\cdot\mathbf{r}_{i}}$,
and then the total momentum distribution can be written as $n\left(\mathbf{q}\right)=\sum_{i=1}^{N}n_{i}\left(\mathbf{q}\right)$.
When two fermions $\left(i,j\right)$ get close while all the other
fermions are far away, we may write the many-body function $\Psi_{3D}$
at $r=\left|{\bf r}_{i}-{\bf r}_{j}\right|\approx0$ as the following
ansatz

\begin{multline}
\Psi_{3D}\left(\mathbf{X},{\bf r}\right)=\left[\frac{\alpha_{0}\left(\mathbf{X}\right)}{r}+\mathcal{B}_{0}\left(\mathbf{X}\right)+\mathcal{C}_{0}\left(\mathbf{X}\right)r\right]\Omega_{0}\left(\hat{{\bf r}}\right)\\
+\left[\frac{\alpha_{1}\left(\mathbf{X}\right)}{r^{2}}+\mathcal{B}_{1}\left(\mathbf{X}\right)+\mathcal{C}_{1}\left(\mathbf{X}\right)r\right]\Omega_{1}\left(\hat{{\bf r}}\right)+O\left(r^{2}\right),\label{eq:lm2}
\end{multline}
where $\alpha_{m}$, $\mathcal{B}_{m}$ and $\mathcal{C}_{m}$ ($m=0,1$)
are all regular functions. Comparing Eq. (\ref{eq:f11}) with (\ref{eq:lm2})
at small $r$, we find

\begin{eqnarray}
\mathcal{B}_{0}\left(\mathbf{X}\right) & = & -\frac{\alpha_{0}}{a_{0}}+\alpha_{1}u\lambda,\\
\mathcal{B}_{1}\left(\mathbf{X}\right) & = & \frac{\alpha_{1}k^{2}}{2}+\alpha_{0}\lambda,\\
\mathcal{C}_{0}\left(\mathbf{X}\right) & = & -\frac{\alpha_{0}k^{2}}{2},\\
\mathcal{C}_{1}\left(\mathbf{X}\right) & = & -\frac{\alpha_{1}}{3a_{1}}+\frac{\alpha_{1}b_{1}k^{2}}{6}+\alpha_{0}v\lambda.
\end{eqnarray}
The asymptotic form of the momentum distribution at large $\mathbf{q}$
but still smaller than $\epsilon^{-1}$ is determined by the asymptotic
behavior at short distance with respect to the two interacting fermions,
then we have

\begin{equation}
\tilde{\Psi}_{i}\left(\mathbf{q}\right)\underset{q\rightarrow\infty}{\approx}\int d\mathbf{r}\Psi_{3D}\left(\mathbf{X},{\bf r}\sim0\right)e^{-i\mathbf{q}\cdot\mathbf{r}}.\label{eq:ns1}
\end{equation}
Using $\nabla^{2}\left(r^{-1}\right)=-4\pi\delta\left(\mathbf{r}\right)$,
we have the identity

\begin{equation}
f(q)\equiv\int d\mathbf{r}\frac{e^{-i\mathbf{q}\cdot\mathbf{r}}}{r}=\frac{4\pi}{q^{2}},
\end{equation}
so that

\begin{eqnarray}
\int d\mathbf{r}\frac{\alpha_{0}\left(\mathbf{X}\right)}{r}\Omega_{0}\left(\hat{{\bf r}}\right)e^{-i\mathbf{q}\cdot\mathbf{r}} & = & \frac{4\pi}{q^{2}}\alpha_{0}\left(\mathbf{X}\right)\Omega_{0}\left(\hat{\mathbf{q}}\right),\label{eq:ds1}\\
\int d\mathbf{r}\mathcal{B}_{0}\left(\mathbf{X}\right)\Omega_{0}\left(\hat{{\bf r}}\right)e^{-i\mathbf{q}\cdot\mathbf{r}} & = & 0,\label{eq:ds2}\\
\int d\mathbf{r}\mathcal{C}_{0}\left(\mathbf{X}\right)r\Omega_{0}\left(\hat{{\bf r}}\right)e^{-i\mathbf{q}\cdot\mathbf{r}} & = & -\frac{8\pi}{q^{4}}\mathcal{C}_{0}\left(\mathbf{X}\right)\Omega_{0}\left(\hat{\mathbf{q}}\right),\label{eq:ds3}\\
\int d\mathbf{r}\frac{\alpha_{1}\left(\mathbf{X}\right)}{r^{2}}\Omega_{1}\left(\hat{{\bf r}}\right)e^{-i\mathbf{q}\cdot\mathbf{r}} & = & -i\frac{4\pi}{q}\alpha_{1}\left(\mathbf{X}\right)\Omega_{1}\left(\hat{\mathbf{q}}\right),\label{eq:ds4}\\
\int d\mathbf{r}\mathcal{B}_{1}\left(\mathbf{X}\right)\Omega_{1}\left(\hat{{\bf r}}\right)e^{-i\mathbf{q}\cdot\mathbf{r}} & = & -i\frac{8\pi}{q^{3}}\mathcal{B}_{1}\left(\mathbf{X}\right)\Omega_{1}\left(\hat{\mathbf{q}}\right),\label{eq:ds5}\\
\int d\mathbf{r}\mathcal{C}_{1}\left(\mathbf{X}\right)r\Omega_{1}\left(\hat{{\bf r}}\right)e^{-i\mathbf{q}\cdot\mathbf{r}} & = & 0.\label{eq:ds6}
\end{eqnarray}
Inserting Eqs. (\ref{eq:ds1})-(\ref{eq:ds6}) into (\ref{eq:ns1}),
and then into Eq. (\ref{eq:lm1}), we find that the total momentum
distribution $n\left(\mathbf{q}\right)$ at large $\mathbf{q}$ takes
the form

\begin{multline}
n_{3D}\left(\mathbf{q}\right)\approx\mathcal{N}\int d\mathbf{X}\frac{32\pi^{2}\alpha_{1}\alpha_{1}^{*}\Omega_{1}\left(\hat{\mathbf{q}}\right)\Omega_{1}^{*}\left(\hat{\mathbf{q}}\right)}{q^{2}}\\
+i\frac{32\pi^{2}\left[\alpha_{0}\alpha_{1}^{*}\Omega_{0}\left(\hat{\mathbf{q}}\right)\Omega_{1}^{*}\left(\hat{\mathbf{q}}\right)-\alpha_{0}^{*}\alpha_{1}\Omega_{0}^{*}\left(\hat{\mathbf{q}}\right)\Omega_{1}\left(\hat{\mathbf{q}}\right)\right]}{q^{3}}\\
+\left[32\pi^{2}\alpha_{0}\alpha_{0}^{*}\Omega_{0}\left(\hat{\mathbf{q}}\right)\Omega_{0}^{*}\left(\hat{\mathbf{q}}\right)+64\pi^{2}k^{2}\alpha_{1}\alpha_{1}^{*}\Omega_{1}\left(\hat{\mathbf{q}}\right)\Omega_{1}^{*}\left(\hat{\mathbf{q}}\right)\right.\\
\left.+64\pi^{2}\lambda\left(\alpha_{0}\alpha_{1}^{*}+\alpha_{0}^{*}\alpha_{1}\right)\Omega_{1}\left(\hat{\mathbf{q}}\right)\Omega_{1}^{*}\left(\hat{\mathbf{q}}\right)\right]\frac{1}{q^{4}}+O\left(q^{-5}\right).\label{eq:tas4-1}
\end{multline}
If we are only interested in the dependence of the momentum distribution
on the amplitude of $\mathbf{q}$, we may integrate over the direction
of $\mathbf{q}$, and then we find all the odd-order terms of $q^{-1}$
vanish. Finally, we obtain

\begin{equation}
n_{3D}\left(q\right)=\frac{C_{a}^{\left(1\right)}}{q^{2}}+\left(C_{a}^{\left(0\right)}+C_{b}^{\left(1\right)}+\lambda\mathcal{P}_{\lambda}\right)\frac{1}{q^{4}}+O\left(q^{-6}\right),\label{eq:lz4-1}
\end{equation}
where the contacts are defined as

\begin{eqnarray}
C_{a}^{\left(m\right)} & = & 32\pi^{2}\mathcal{I}_{a}^{\left(m\right)},\,\left(m=0,1\right),\label{eq:lp8}\\
C_{b}^{\left(1\right)} & = & \frac{64\pi^{2}M}{\hbar^{2}}\mathcal{E}_{1},\label{eq:lp9}\\
\mathcal{P}_{\lambda} & = & 64\pi^{2}\mathcal{I}_{\lambda}.\label{eq:lp10}
\end{eqnarray}
With these definitions in hands, the adiabatic energy relations (\ref{eq:taz9})-(\ref{eq:taz16})
can alternatively be written as

\begin{eqnarray}
\frac{\partial E}{\partial a_{0}^{-1}} & = & -\frac{\hbar^{2}C_{a}^{\left(0\right)}}{32\pi^{2}M}+\lambda\frac{\hbar^{2}\mathcal{P}_{\lambda}}{64\pi^{2}M},\label{eq:taz9-1}\\
\frac{\partial E}{\partial a_{1}^{-1}} & = & -\frac{\hbar^{2}C_{a}^{\left(1\right)}}{32\pi^{2}M},\label{eq:taz11-1}\\
\frac{\partial E}{\partial b_{1}} & = & \frac{\hbar^{2}C_{b}^{\left(1\right)}}{128\pi^{2}M},\label{eq:taz12-1}\\
\frac{\partial E}{\partial u} & = & \lambda\left[\frac{\hbar^{2}\mathcal{P}_{\lambda}}{128\pi^{2}M}-\lambda\frac{\hbar^{2}C_{a}^{\left(1\right)}}{16\pi^{2}M}\right],\label{eq:taz15-1}\\
\frac{\partial E}{\partial v} & = & \frac{3\lambda\hbar^{2}\mathcal{P}_{\lambda}}{128\pi^{2}M}.\label{eq:taz16-1}
\end{eqnarray}
In the absence of SO coupling, Eqs. (\ref{eq:taz9-1}), (\ref{eq:taz11-1})
and (\ref{eq:taz12-1}) simply reduce to the ordinary form of the
adiabatic energy relations for $s$- and $p$-wave interactions \cite{Tan2008L,Peng2016L},
with respect to the scattering length (or volume) as well as effective
range. We should note that for the $s$-wave interaction, there is
a difference of the factor $8\pi$ from the well-known form of adiabatic
energy relations. This is because we include the spherical harmonics
$Y_{00}\left(\mathbf{\hat{r}}\right)=1/\sqrt{4\pi}$ in the $s$-partial
wave function. Besides, an additional factor $1/2$ is introduced
in order to keep the definition of contacts consistent with those
in the tail of the momentum distribution at large $\mathbf{q}$. In
the presence of SO coupling, two additional adiabatic energy relations
appear, i.e., Eqs. (\ref{eq:taz15-1}) and (\ref{eq:taz16-1}), and
a new contact $\mathcal{P}_{\lambda}$ is introduced.

\subsection{The high-frequency tail of the rf spectroscopy}

Next, we discuss the asymptotic behavior of the rf spectroscopy at
high frequency. The basic ideal of the rf transition is as follows.
For an atomic Fermi gas with two hyperfine states, denoted as $\left|\uparrow\right\rangle $
and $\left|\downarrow\right\rangle $, the rf field drives transitions
between one of the hyperfine states ( i.e. $\left|\downarrow\right\rangle $)
and an empty hyperfine state $\left|3\right\rangle $ with a bare
atomic hyperfine energy difference $\hbar\omega_{3\downarrow}$ due
to the magnetic field splitting \cite{Hu2012R,Peng2012M}. The universal
scaling behavior at high frequency of the rf response of the system
is governed by contacts. In this subsection, we are going to show
how the contacts defined by the adiabatic energy relations characterize
such high-frequency scalings of the rf transition in 3D Fermi gases
with 3D SO coupling. Here, we will present a two-body derivation first,
which may avoid complicated notations as much as possible, and the
results can easily be generalized to many-body systems later. The
rf field driving the spin-down particle to the state $\left|3\right\rangle $
is described by

\begin{equation}
\mathcal{H}_{rf}=\gamma_{rf}\underset{{\bf k}}{\sum}\left(e^{-i\omega t}c_{3\mathbf{k}}^{\dagger}c_{\downarrow\mathbf{k}}+e^{i\omega t}c_{\downarrow\mathbf{k}}^{\dagger}c_{3\mathbf{k}}\right),\label{eq:rf2}
\end{equation}
where $\gamma_{rf}$ is the strength of the rf drive, $\omega$ is
the rf frequency, and $c_{\sigma\mathbf{k}}^{\dagger}$ and $c_{\sigma{\bf k}}$
are respectively the creation and annihilation operators for fermions
with the momentum $\mathbf{k}$ in the spin states $\left|\sigma\right\rangle $.

For any two-body state $\left|\Psi_{2b}\right\rangle $, we may write
it in the momentum space as

\begin{equation}
\left|\Psi_{2b}\right\rangle =\sum_{\sigma_{1}\sigma_{2}}\underset{\mathbf{k}_{1}\mathbf{k}_{2}}{\sum}\tilde{\phi}_{\sigma_{1}\sigma_{2}}\left({\bf k}_{1},{\bf k}_{2}\right)c_{\sigma_{1}{\bf k}_{1}}^{\dagger}c_{\sigma_{2}{\bf k}_{2}}^{\dagger}\left|0\right\rangle ,\label{eq:2b3-1}
\end{equation}
where $\tilde{\phi}_{\sigma_{1}\sigma_{2}}\left(\mathbf{k_{\mathrm{1}}},\mathbf{k}_{2}\right)$
is the Fourier transform of $\phi_{\sigma_{1}\sigma_{2}}\left({\bf r}_{1},{\bf r}_{2}\right)\equiv\left\langle {\bf r}_{1},{\bf r}_{2};\sigma_{1},\sigma_{2}|\Psi_{2b}\right\rangle $,
i.e.,
\begin{equation}
\tilde{\phi}_{\sigma_{1}\sigma_{2}}\left(\mathbf{k_{\mathrm{1}}},\mathbf{k}_{2}\right)=\int d{\bf r}_{1}d{\bf r}_{2}\phi_{\sigma_{1}\sigma_{2}}\left({\bf r}_{1},{\bf r}_{2}\right)e^{-i{\bf k}_{1}\cdot{\bf r}_{1}}e^{-i{\bf k}_{2}\cdot{\bf r}_{2}},\label{eq:rf3}
\end{equation}
and $\sigma_{i}=\uparrow,\downarrow$ denotes the spin of the $i$th
particle. The specific form of $\tilde{\phi}_{\sigma_{1}\sigma_{2}}\left(\mathbf{k_{\mathrm{1}}},\mathbf{k}_{2}\right)$
can easily be obtained by using that of the two-body wave function
$\left\langle {\bf r}_{1},{\bf r}_{2};\sigma_{1},\sigma_{2}|\Psi_{2b}\right\rangle $
in the coordinate space, i.e., Eq.(\ref{eq:2bw}). Acting Eq.(\ref{eq:rf2})
onto (\ref{eq:2b3-1}), we obtain the two-body wave function after
the rf transition,
\begin{multline}
\mathcal{H}_{rf}\left|\Psi_{2b}\right\rangle =\gamma_{rf}e^{-i\omega t}\times\\
\sum_{{\bf k}_{1}{\bf k}_{2}}\left[\tilde{\phi}_{\downarrow\uparrow}\left(\mathbf{k_{\mathrm{1}}},\mathbf{k}_{2}\right)c_{\mathbf{3k_{1}}}^{\dagger}c_{\uparrow\mathbf{k_{\mathrm{2}}}}^{\dagger}-\tilde{\phi}_{\uparrow\downarrow}\left(\mathbf{k_{\mathrm{1}}},\mathbf{k}_{2}\right)c_{\mathbf{3k_{\mathrm{2}}}}^{\dagger}c_{\uparrow\mathbf{k_{\mathrm{1}}}}^{\dagger}\right.\\
\left.+\tilde{\phi}_{\downarrow\downarrow}\left(\mathbf{\mathbf{k_{\mathrm{1}}},\mathbf{k}_{2}}\right)\left(c_{3\mathbf{k_{\mathrm{1}}}}^{\dagger}c_{\downarrow\mathbf{k_{\mathrm{2}}}}^{\dagger}-c_{3\mathbf{k_{\mathrm{2}}}}^{\dagger}c_{\downarrow\mathbf{k_{\mathrm{1}}}}^{\dagger}\right)\right]\left|0\right\rangle .\label{eq:rf4}
\end{multline}
The physical meaning of Eq.(\ref{eq:rf4}) is apparent: after the
rf transition, the atom with initial spin state $\left|\downarrow\right\rangle $
is driven to the empty spin state $\left|3\right\rangle $, while
the other one remains in the spin state $\left|\uparrow\right\rangle $.
Therefore, there are totally four possible final two-body states with,
respectively, possibilities of $\left|\tilde{\phi}_{\downarrow\uparrow}\right|^{2}$,
$\left|\tilde{\phi}_{\uparrow\downarrow}\right|^{2}$, $\left|\tilde{\phi}_{\downarrow\downarrow}\right|^{2}$,
and $\left|\tilde{\phi}_{\downarrow\downarrow}\right|^{2}$. Taking
all these final states into account, and according to the Fermi's
golden rule \cite{Peng2019U}, the two-body rf transition rate is
therefore given by the Franck-Condon factor,

\begin{multline}
\Gamma_{2}\left(\omega\right)=\frac{2\pi\gamma_{rf}^{2}}{\hbar}\times\\
\underset{{\bf k}_{1}{\bf k}_{2}}{\sum}\left(\left|\tilde{\phi}_{\downarrow\uparrow}\right|^{2}+\left|\tilde{\phi}_{\uparrow\downarrow}\right|^{2}+2\left|\tilde{\phi}_{\downarrow\downarrow}\right|^{2}\right)\delta\left(\hbar\omega-\triangle E\right),\label{eq:tr1}
\end{multline}
where $\triangle E$ is the energy difference between the final and
initial states, and takes the form of
\begin{equation}
\Delta E=\frac{\hbar^{2}k^{2}}{M}-\frac{\hbar^{2}q^{2}}{M}+\hbar\omega_{3\downarrow},
\end{equation}
where $\mathbf{k}=\left(\mathbf{k}_{1}-\mathbf{k}_{2}\right)/2$,
$\hbar^{2}q^{2}/M$ is the relative energy of two fermions in the
initial state, and $\omega_{3\downarrow}\equiv\omega_{3}-\omega_{\downarrow}$
is the bare hyperfine splitting between the spin states $\left|3\right\rangle $
and $\left|\downarrow\right\rangle $, and can be set to $0$ without
loss of generality. Now, we are interested in the asymptotic form
of $\Gamma_{2}\left(\omega\right)$ at large $\omega$ but still small
compared to $\hbar/M\epsilon^{2}$, which is determined by the short-range
behavior when two fermions get as close as $\epsilon$. Combining
Eqs.(\ref{eq:rf3}) and (\ref{eq:tr1}), as well as the asymptotic
form of the two-body wave function (\ref{eq:f11}) at ${\bf r}={\bf r}_{1}-{\bf r}_{2}\sim0$,
we finally obtain the asymptotic behavior of the rf response of 3D
SO-coupled Fermi gases at large $\omega$,

\begin{multline}
\Gamma_{2}\left(\omega\right)=\frac{M\gamma_{rf}^{2}}{16\pi^{2}\hbar^{3}}\left[\frac{c_{a}^{\left(1\right)}}{\left(M\omega/\hbar\right)^{1/2}}+\right.\\
\left.\frac{c_{a}^{\left(0\right)}+3c_{b}^{\left(1\right)}/4+\lambda p_{\lambda}}{\left(M\omega/\hbar\right)^{3/2}}\right],\label{eq:tr4}
\end{multline}
where $c_{a}^{\left(0\right)}$, $c_{a}^{\left(1\right)}$, $c_{b}^{\left(1\right)}$
and $p_{\lambda}$ are contacts for a two-body system with $\mathcal{N}=1$
in the definitions (\ref{eq:lp8})-(\ref{eq:lp10}). 

For many-body systems, all possible $\mathcal{N}=N\left(N-1\right)/2$
pairs may contribute to the high-frequency tail of the rf spectroscopy,
while high-order contributions from more than two fermions are ignored.
Then we can generalize the above two-body picture to many-body systems
by simply redefining the constant $\mathcal{N}$ into the contacts,
and then obtain

\begin{eqnarray}
\Gamma_{N}\left(\omega\right) & = & \frac{M\gamma_{rf}^{2}}{16\pi^{2}\hbar^{3}}\left[\frac{C_{a}^{\left(1\right)}}{\left(M\omega/\hbar\right)^{1/2}}+\right.\nonumber \\
 &  & \left.\qquad\frac{C_{a}^{\left(0\right)}+3C_{b}^{\left(1\right)}/4+\lambda\mathcal{P}_{\lambda}}{\left(M\omega/\hbar\right)^{3/2}}\right],\label{eq:tr4-1}
\end{eqnarray}
where $C_{a}^{\left(0\right)}$, $C_{a}^{\left(1\right)}$, $C_{b}^{\left(1\right)}$
and $\mathcal{P}_{\lambda}$ are corresponding contacts for many-body
systems. In the absence of SO coupling, Eq. (\ref{eq:tr4-1}) simply
reduces to the ordinary asymptotic behaviors of the rf response for
$s$- and $p$-wave interactions, respectively \cite{Yu2015U,Braaten2010S}.

\subsection{Pair correlation function at short distances}

The pair correlation function $g_{2}\left(\mathbf{s}_{1},\mathbf{s}_{2}\right)$
gives the probability of finding two fermions with one at position
$\mathbf{s}_{1}$ and the other one at position $\mathbf{s}_{2}$
simultaneously, i.e., $g_{2}\left({\bf s}_{1},{\bf s}_{2}\right)\equiv\left\langle \hat{\rho}\left({\bf s}_{1}\right)\hat{\rho}\left({\bf s}_{2}\right)\right\rangle $,
where $\hat{\rho}\left({\bf s}\right)=\sum_{i}\delta\left({\bf s}-{\bf r}_{i}\right)$
is the density operator at the position ${\bf s}$. For a pure many-body
state $\left|\Psi\right\rangle $ of $N$ fermions, we have \cite{Peng2019U}

\begin{eqnarray}
g_{2}\left(\mathbf{s}_{1},\mathbf{s}_{2}\right) & = & \int d\mathbf{r}_{1}d\mathbf{r}_{2}\cdots d\mathbf{r}_{N}\left\langle \Psi\right|\hat{\rho}\left(\mathbf{s}_{1}\right)\hat{\rho}\left(\mathbf{s}_{2}\right)\left|\Psi\right\rangle \nonumber \\
 & = & N\left(N-1\right)\int d\mathbf{X}^{\prime}\left|\Psi\left(\mathbf{X},\mathbf{r}\right)\right|^{2},\label{eq:pc1}
\end{eqnarray}
where $\mathbf{r}=\mathbf{s}_{1}-\mathbf{s}_{2}$ is relative coordinates
of the pair fermions at positions $\mathbf{s}_{1}$ and $\mathbf{s}_{2}$,
and $\mathbf{X}^{\prime}$ denotes the degrees of freedom of all the
other fermions. If we further integrate over the c.m. coordinate of
the pair, we can define the spatially integrated pair correlation
function as

\begin{equation}
G_{2}\left(\mathbf{r}\right)\equiv N\left(N-1\right)\int d\mathbf{X}\left|\Psi\left(\mathbf{X},\mathbf{r}\right)\right|^{2},\label{eq:pc2}
\end{equation}
and ${\bf X}$ includes the c.m. coordinate ${\bf R}=\left({\bf s}_{1}+{\bf s}_{2}\right)/2$
of the pair besides ${\bf X}^{\prime}$. Inserting the short-range
form of many-body wave functions for SO coupled Fermi gases, i.e.
Eq.(\ref{eq:f11}) into Eq. (\ref{eq:pc2}), we find

\begin{align}
G_{2}\left(\mathbf{r}\right) & \approx N\left(N-1\right)\int d\mathbf{X}\left\{ \mathbf{\frac{\alpha_{\mathrm{1}}\alpha_{\mathrm{1}}^{*}\mathrm{\Omega_{\mathrm{1}}\Omega_{\mathrm{1}}^{*}}}{\mathrm{\mathit{r}^{4}}}}\right.\nonumber \\
 & +\mathbf{\frac{\alpha_{\mathrm{0}}^{*}\alpha_{\mathrm{1}}\mathrm{\Omega_{\mathrm{0}}^{*}\Omega_{\mathrm{1}}}+\alpha_{\mathrm{0}}\alpha_{\mathrm{1}}^{*}\mathrm{\Omega_{\mathrm{0}}\Omega_{\mathrm{1}}^{*}}}{\mathrm{\mathit{r}^{3}}}}+\left[\alpha_{\mathrm{0}}\alpha_{\mathrm{0}}^{*}\Omega_{\mathrm{0}}\Omega_{\mathrm{0}}^{*}\right.\nonumber \\
 & +k^{2}\alpha_{\mathrm{1}}\alpha_{\mathrm{1}}^{*}\Omega_{\mathrm{1}}\Omega_{\mathrm{1}}^{*}+\lambda\left(\alpha_{\mathrm{0}}^{*}\alpha_{\mathrm{1}}+\alpha_{\mathrm{0}}\alpha_{\mathrm{1}}^{*}\right)\Omega_{\mathrm{1}}^{*}\Omega_{\mathrm{1}}\nonumber \\
 & +\lambda u\alpha_{\mathrm{1}}^{*}\alpha_{\mathrm{1}}\left(\Omega_{\mathrm{0}}^{*}\Omega_{\mathrm{1}}+\Omega_{\mathrm{0}}\Omega_{\mathrm{1}}^{*}\right)\nonumber \\
 & \left.\left.-\frac{\alpha_{\mathrm{0}}^{*}\alpha_{\mathrm{1}}\Omega_{\mathrm{0}}^{*}\Omega_{\mathrm{1}}+\alpha_{\mathrm{0}}\alpha_{\mathrm{1}}^{*}\Omega_{\mathrm{0}}\Omega_{\mathrm{1}}^{*}}{a_{0}}\right]\frac{1}{r^{2}}+O\left(r^{-1}\right)\right\} .
\end{align}
Further, if we are only care about the dependence of $G_{2}\left(\mathbf{r}\right)$
on the amplitude of $r=\left|\mathbf{r}\right|$, we can integrate
over the direction of $\mathbf{r}$, and use the definitions of contacts
(\ref{eq:lp8})-(\ref{eq:lp10}), then it yields

\begin{multline}
G_{2}\left(r\right)\approx\frac{1}{16\pi^{2}}\left[\frac{C_{a}^{\left(1\right)}}{r^{4}}+\left(C_{a}^{\left(0\right)}+\frac{C_{b}^{\left(1\right)}}{2}+\lambda\frac{\mathcal{P}_{\lambda}}{2}\right)\frac{1}{r^{2}}\right.\\
\left.+\left(-\frac{2C_{a}^{\left(0\right)}}{a_{0}}-\frac{2C_{a}^{\left(1\right)}}{3a_{1}}+\frac{b_{1}C_{b}^{\left(1\right)}}{6}+\lambda\left(u+v\right)\frac{\mathcal{P}_{\lambda}}{2}\right)\frac{1}{r}\right].\label{eq:mg1}
\end{multline}
which reduces to the results in the absence of the SO coupling for
$s$- and $p$-wave interactions, respectively\cite{Tan2008E,Yu2015U,Kuhnle2010U,Gandolfi2011B,Sakumichi2014L}.

\subsection{Grand canonical potential and pressure relation}

The adiabatic energy relations as well as the large-momentum distribution
we obtained is valid for any pure energy eigenstate. Therefore, they
should still hold for any incoherent mixed state statistically at
finite temperature. Then the energy $E$ and contacts then become
their statistical average values. Now, let us look at the grand thermodynamic
potential $\mathcal{J}$ for a homogeneous system, which is defined
as \cite{Landau2007S}

\begin{equation}
\mathcal{J}\equiv-PV=E-TS-\mu N,\label{eq:g1}
\end{equation}
where $P$, $V$, $T$, $S$, $\mu$, $N$ are, respectively, the
pressure, volume, temperature, entropy, chemical potential, and total
particle number. The grand canonical potential $\mathcal{J}$ is the
function of $V$, $T$, $S$, and takes the following differential
form

\begin{equation}
d\mathcal{J}=-PdV-SdT-Nd\mu.\label{eq:g2}
\end{equation}
For the two-body microscopic parameters, we may evaluate their dimensions
as $a_{0}\sim\mathrm{Length^{1}}$, $a_{1}\sim\mathrm{Length^{3}}$,
$b_{1}\sim\mathrm{Length^{-1}}$, $u\sim\mathrm{Length^{-1}}$, and
$v\sim\mathrm{Length^{-1}}$. Therefore, there are basically following
energy scales in the grand thermodynamic potential, i.e., $k_{B}T$,
$\mu$, $\hbar^{2}/MV^{2/3}$, $\hbar^{2}/Ma_{0}^{2}$, $\hbar^{2}/Ma_{1}^{2/3}$,
$\hbar^{2}b_{1}^{2}/M$, $\hbar^{2}u^{2}/M$, $\hbar^{2}v^{2}/M$.
Then we may express the thermodynamic potential $\mathcal{J}$ in
the terms of a dimensionless function $\mathcal{\bar{J}}$ as \cite{Barth2011T,Braaten2008U}

\begin{multline}
\mathcal{J}\left(V,T,\mu,a_{0},a_{1},b_{1},u,v\right)\\
=k_{B}T\bar{\mathcal{J}}\left(\frac{\hbar^{2}/MV^{2/3}}{k_{B}T},\frac{\mu}{k_{B}T},\frac{\hbar^{2}/Ma_{0}^{2}}{k_{B}T},\right.\\
\left.\frac{\hbar^{2}/Ma_{1}^{2/3}}{k_{B}T},\frac{\hbar^{2}b_{1}^{2}/M}{k_{B}T},\frac{\hbar^{2}u^{2}/M}{k_{B}T},\frac{\hbar^{2}v^{2}/M}{k_{B}T}\right).\label{eq:g3}
\end{multline}
Consequently, one can deduce the simple scaling law

\begin{multline}
\mathcal{J}\left(\gamma^{-3/2}V,\gamma T,\gamma\mu,\gamma^{-1/2}a_{0},\gamma^{-3/2}a_{1},\gamma^{1/2}b_{1},\gamma^{1/2}u,\gamma^{1/2}v\right)\\
=\gamma\mathcal{J}\left(V,T,\mu,a_{0},a_{1},b_{1},u,v\right).\label{eq:g4}
\end{multline}
The derivative of Eq.(\ref{eq:g4}) with respect to $\gamma$ at $\gamma=1$
simply yields

\begin{multline}
\left(-\frac{3V}{2}\frac{\partial}{\partial V}+T\frac{\partial}{\partial T}+\mu\frac{\partial}{\partial\mu}-\frac{a_{0}}{2}\frac{\partial}{\partial a_{0}}\right.\\
\left.-\frac{3a_{1}}{2}\frac{\partial}{\partial a_{1}}+\frac{b_{1}}{2}\frac{\partial}{\partial b_{1}}+\frac{u}{2}\frac{\partial}{\partial u}+\frac{v}{2}\frac{\partial}{\partial v}\right)\mathcal{J}=\mathcal{J},\label{eq:g4-1}
\end{multline}
where all the partial derivatives are to be understood as leaving
all other system variables constant. Since

\begin{equation}
\mathcal{J}-T\frac{\partial\mathcal{J}}{\partial T}-\mu\frac{\partial\mathcal{J}}{\partial\mu}=\mathcal{J}+TS+\mu N=E,\label{eq:g25}
\end{equation}
and the variation of the grand thermodynamic potential $\delta\mathcal{J}$
with respect to the two-body parameters at fixed volume $V$, temperature
$T$ and chemical potential $\mu$ is equal to that of the energy
$\delta E$ at fixed volume $V$, entropy $S$ and particle number
$N$, i.e., $\left(\delta\mathcal{J}\right)_{V,T,\mu}=\left(\delta E\right)_{V,S,N}$
and $V\frac{\partial\mathcal{J}}{\partial V}=\mathcal{J}$, we easily
obtain from Eqs.(\ref{eq:g4-1}) and (\ref{eq:g25})
\begin{multline}
-\frac{3}{2}\mathcal{J}-\frac{a_{0}}{2}\frac{\partial E}{\partial a_{0}}-\frac{3a_{1}}{2}\frac{\partial E}{\partial a_{1}}+\frac{b_{1}}{2}\frac{\partial E}{\partial b_{1}}+\frac{u}{2}\frac{\partial E}{\partial u}+\frac{v}{2}\frac{\partial E}{\partial v}=E,\\
\label{eq:g27}
\end{multline}
Further by using adiabatic energy relations, Eq.(\ref{eq:g27}) becomes

\begin{multline}
\mathcal{J}=-\frac{2}{3}E-\frac{\hbar^{2}}{96\pi^{2}Ma_{0}}\left(C_{a}^{\left(0\right)}-\lambda\frac{\mathcal{P}_{\lambda}}{2}\right)\\
-\frac{\hbar^{2}C_{a}^{\left(1\right)}}{32\pi^{2}Ma_{1}}+\frac{\hbar^{2}b_{1}C_{b}^{\left(1\right)}}{384\pi^{2}M}\\
-\frac{\lambda u\hbar^{2}}{48\pi^{2}M}\left(\lambda C_{a}^{\left(1\right)}-\frac{\mathcal{P}_{\lambda}}{8}\right)+\frac{\lambda v\mathcal{P}_{\lambda}\hbar^{2}}{128\pi^{2}M},\label{eq:g1-1}
\end{multline}
or the pressure relation by dividing both sides of Eq. (\ref{eq:g1-1})
by $-V$, which respectively reduces to the well-known results in
the absence of the spin-orbit coupling 

\begin{equation}
P=\frac{2E}{3V}+\frac{\hbar^{2}C_{a}^{\left(0\right)}}{96\pi^{2}MVa_{0}}\label{eq:g31-1}
\end{equation}
for $s$-wave interactions, which is consistent with the result of
Ref.\cite{Tan2008G,Sakumichi2014L,Iskin2011I}, and 

\begin{equation}
P=\frac{2E}{3V}+\frac{\hbar^{2}C_{a}^{\left(1\right)}}{32\pi^{2}MVa_{1}}-\frac{b_{1}\hbar^{2}C_{b}^{\left(1\right)}}{384\pi^{2}MV}\label{eq:g31-2}
\end{equation}
for $p$-wave interactions, which is consistent with the result of
Ref.\cite{Yu2015U}.

\section{Universal relations in 2D systems with Rashba SO coupling}

The derivation of the universal relations for 3D Fermi gases with
3D SO coupling can directly be generalized to those for 2D systems
with 2D SO coupling. In this section, with the short-range form of
the two-body wave function for 2D systems with 2D SO coupling in hands,
i.e., Eq.(\ref{eq:u2d}), we are going to discuss Tan's universal
relations for 2D Fermi gases with 2D SO coupling , by taking into
account only two-body correlations.

\subsection{Adiabatic energy relations}

Let us consider how the energy of the SO-coupled system varies with
the two-body interaction in 2D systems with 2D SO coupling. The two
wave functions of a many-body system $\Psi\left(\mathbf{r}\right)$
and $\Psi^{\prime}\left(\mathbf{r}\right)$, corresponding to different
interatomic interaction strengths, satisfy the Schr\"{o}dinger equation
with different energies, i.e. formally as Eqs. (\ref{eq:tas2}) and
(\ref{eq:tas3}). Analogously, by subtracting $[\ref{eq:tas3}]^{*}\times\Psi$
from $\Psi^{\prime*}\times\left[\ref{eq:tas2}\right]$, and integrating
over the domain $\mathcal{D}_{\epsilon}$, the set of all configurations
$\left({\bf r}_{i},{\bf r}_{j}\right)$ in which $r=\left|{\bf r}_{i}-{\bf r}_{j}\right|>\epsilon$,
we obtain

\begin{multline}
\left(E-E^{\prime}\right)\int_{\mathcal{D}_{\epsilon}}\stackrel[i=1]{N}{\prod}d\mathbf{r}_{i}\Psi^{\prime*}\Psi=\\
-\frac{\hbar^{2}}{M}\mathcal{N}\int_{r>\epsilon}d\mathbf{X}d\mathbf{r}\left[\Psi^{\prime*}\nabla_{\mathbf{r}}^{2}\Psi-\left(\nabla_{\mathbf{r}}^{2}\Psi^{\prime*}\right)\Psi\right]\\
+\frac{\hbar^{2}\lambda}{M}\mathcal{N}\int_{r>\epsilon}d\mathbf{X}d\mathbf{r}\left[\Psi^{\prime*}\left(\hat{Q}\Psi\right)-\left(\hat{Q}\Psi^{\prime}\right)^{*}\Psi\right],\label{eq:2da1}
\end{multline}
where $\mathcal{N}=N\left(N-1\right)/2$ is again the number of all
the possible ways to pair atom. Using the Gauss\textquoteright{} theorem,
the first term on the right-hand side (RHS) can be written as 

\begin{align}
 & -\frac{\hbar^{2}}{M}\mathcal{N}\int_{r>\epsilon}d\mathbf{X}d\mathbf{r}\left[\Psi^{\prime*}\nabla_{\mathbf{r}}^{2}\Psi-\left(\nabla_{\mathbf{r}}^{2}\Psi^{\prime*}\right)\Psi\right]\nonumber \\
= & -\frac{\hbar^{2}}{M}\mathcal{N}\oint_{r=\epsilon}\left[\Psi^{\prime*}\nabla_{\mathbf{r}}\Psi-\left(\nabla_{\mathbf{r}}\Psi^{\prime*}\right)\Psi\right]\cdot\hat{\mathbf{n}}d\mathcal{S},\nonumber \\
= & \frac{\hbar^{2}\epsilon}{M}\mathcal{N}\int d\mathbf{X}\underset{m=0,\pm1}{\sum}\left(\psi_{m}^{\prime*}\frac{\partial}{\partial r}\psi_{m}-\psi_{m}\frac{\partial}{\partial r}\psi_{m}^{\prime*}\right)_{r=\epsilon},\label{eq:2da2}
\end{align}
where $\mathcal{S}$ is the boundary of $\mathcal{D}_{\epsilon}$
that the distance between the two fermions in the pair $\left(i,j\right)$
is $\epsilon$, $\hat{\mathbf{n}}$ is the direction normal to $\mathcal{S}$,
but is opposite to the radial direction, and $\psi_{0}$ ($\psi_{\pm1}$)
is the $s$-wave ($p$-wave) component of the two-body wave function
as defined in Eq.(\ref{eq:2Dwf}). Since
\begin{multline}
\hat{Q}\left(\mathbf{r}\right)\Psi=\underset{m=\pm1}{\sum}\left[-\frac{\sqrt{2}}{r}\frac{\partial}{\partial r}\left(r\psi_{m}\right)\Omega_{0}\left(\mathbf{\hat{r}}\right)\right.\\
\left.+\sqrt{2}\frac{\partial\psi_{0}}{\partial r}\Omega_{m}\left(\mathbf{\hat{r}}\right)\right],
\end{multline}
we find that the second term on the right-hand side (RHS) of Eq. (\ref{eq:2da1})
can be written as 

\begin{multline}
\frac{\hbar^{2}\lambda}{M}\mathcal{N}\int_{r>\epsilon}d\mathbf{X}d\mathbf{r}\left[\Psi^{\prime*}\left(\hat{Q}\left(\mathbf{r}\right)\Psi\right)-\left(\hat{Q}\left(\mathbf{r}\right)\Psi^{\prime}\right)^{*}\Psi\right]\\
=\frac{\sqrt{2}\lambda\hbar^{2}\epsilon}{M}\mathcal{N}\int d\mathbf{X}\underset{m=\pm1}{\sum}\left(\psi_{0}^{\prime*}\psi_{m}-\psi_{m}^{\prime*}\psi_{0}\right)_{r=\epsilon}.\label{eq:2da3}
\end{multline}
Combining Eqs.(\ref{eq:2da1}), (\ref{eq:2da2}) and (\ref{eq:2da3}),
we have

\begin{multline}
\left(E-E^{\prime}\right)\int_{\mathcal{D_{\epsilon}}}\stackrel[i=1]{N}{\prod}d\mathbf{r}_{i}\Psi^{\prime*}\Psi\\
=\frac{\hbar^{2}\epsilon}{M}\mathcal{N}\int d\mathbf{X}\underset{m=0,\pm1}{\sum}\left(\psi_{m}^{\prime*}\frac{\partial}{\partial r}\psi_{m}-\psi_{m}\frac{\partial}{\partial r}\psi_{m}^{\prime*}\right)_{r=\epsilon}\\
+\frac{\sqrt{2}\lambda\hbar^{2}\epsilon}{M}\mathcal{N}\int d\mathbf{X}\underset{m=\pm1}{\sum}\left(\psi_{0}^{\prime*}\psi_{m}-\psi_{m}^{\prime*}\psi_{0}\right)_{r=\epsilon}.\label{eq:2da4}
\end{multline}
Inserting the asymptotic form of the many-body wave function Eq.(\ref{eq:u2d})
into Eq.(\ref{eq:2da4}), and letting $E^{\prime}\rightarrow E$ and
$\Psi^{\prime}\rightarrow\Psi$, we arrive at

\begin{multline}
\delta E\cdot\int_{\mathcal{D_{\mathrm{\epsilon}}}}\stackrel[i=1]{N}{\prod}d\mathbf{r}_{i}\left|\Psi\right|^{2}\\
=\frac{\hbar^{2}}{M}\left(\mathcal{I}_{a}^{\left(0\right)}+\underset{m=\pm1}{\sum}\frac{\sqrt{2}}{2}\lambda\mathcal{I}_{\lambda}^{(m)}\right)\delta\ln a_{0}\\
+\underset{m=\pm1}{\sum}\left\{ -\frac{\pi\hbar^{2}\mathcal{I}_{a}^{\left(m\right)}}{2M}\delta a_{1}^{-1}+\left(\mathcal{E}_{m}-\frac{\lambda\hbar^{2}\mathcal{I_{\lambda}^{\left(\mathit{\mathit{m}}\right)}}}{\sqrt{2}M}\right)\delta\ln b_{1}\right.\\
-\frac{\hbar^{2}}{M}\left[\left(\sqrt{2}\lambda^{2}\mathcal{I}_{a}^{\left(\mathit{m}\right)}+\frac{\lambda}{2}\mathcal{I_{\lambda}^{\left(\mathit{m}\right)}}\right)+\frac{\sqrt{2}\lambda^{2}}{2}\mathcal{I}_{p}\right]\delta u\\
\left.+\frac{\lambda\hbar^{2}}{M}\mathcal{I}_{\lambda}^{\left(\mathit{\mathrm{\mathit{m}}}\right)}\delta v-\left(\ln\frac{\epsilon}{2b_{1}}+\gamma\right)\mathcal{I}_{a}^{\left(\mathit{m}\right)}\delta E\right\} ,\label{eq:2da5}
\end{multline}
where
\begin{eqnarray}
\mathcal{I}_{a}^{\left(m\right)} & = & \mathcal{N}\int d\mathbf{X}\left|\alpha_{m}\right|^{2},\\
\mathcal{E}_{m} & = & \mathcal{N}\int d\mathbf{X}\alpha_{m}^{*}\left(E-\hat{T}\right)\alpha_{m}
\end{eqnarray}
for $m=0,\pm1$, 
\begin{eqnarray}
\mathcal{I}_{\lambda}^{\left(\pm1\right)} & = & \mathcal{N}\int d\mathbf{X}\alpha_{0}^{*}\alpha_{\pm1}+c.c,\\
\mathcal{E}_{\lambda}^{\left(\pm1\right)} & = & \mathcal{N}\int d\mathbf{X}\alpha_{0}^{*}\left(E-\hat{T}\right)\alpha_{\pm1}+c.c,\\
\mathcal{I}_{p} & = & \mathcal{N}\int d\mathbf{X}\alpha_{-1}^{*}\alpha_{1}+c.c.,
\end{eqnarray}
and $\hat{T}(\mathbf{X})$ is the kinetic operator including the c.m.
motion of the pair as well as those of all the rest fermions. Using
the normalization of the wave function (see appendix B)

\begin{multline}
\int_{\mathcal{D_{\epsilon}}}\stackrel[i=1]{N}{\prod}d\mathbf{r}_{i}\left|\Psi\right|^{2}=1-\underset{m=\pm1}{\sum}\left(\ln\frac{\epsilon}{2b_{1}}+\gamma\right)\mathcal{I}_{a}^{\left(\mathit{m}\right)},
\end{multline}
we can further simplify Eq. (\ref{eq:2da5}) as

\begin{multline}
\delta E=\frac{\hbar^{2}}{M}\left(\mathcal{I}_{a}^{\left(0\right)}+\underset{m=\pm1}{\sum}\lambda\frac{\mathcal{I}_{\lambda}^{\left(\mathrm{\mathit{m}}\right)}}{\sqrt{2}}\right)\delta\ln a_{0}\\
+\underset{m=\pm1}{\sum}\left\{ -\frac{\pi\hbar^{2}\mathcal{I}_{a}^{\left(m\right)}}{2M}\delta a_{1}^{-1}+\left(\mathcal{E}_{m}-\lambda\frac{\hbar^{2}\mathcal{I_{\lambda}^{\left(\mathit{\mathit{m}}\right)}}}{\sqrt{2}M}\right)\delta\ln b_{1}\right.\\
-\lambda\frac{\hbar^{2}}{M}\left[\sqrt{2}\lambda\mathcal{I}_{a}^{\left(\mathit{m}\right)}+\frac{\mathcal{I_{\lambda}^{\left(\mathit{m}\right)}}}{2}+\lambda\frac{\mathcal{I}_{p}}{\sqrt{2}}\right]\delta u\left.+\frac{\lambda\hbar^{2}}{M}\mathcal{I}_{\lambda}^{\left(\mathit{\mathrm{\mathit{m}}}\right)}\delta v\right\} ,\\
\end{multline}
which characterizes how the energy of a 2D system with 2D SO coupling
varies as the scattering parameters adiabatically change, and yields
the following set of adiabatic energy relations

\begin{eqnarray}
\frac{\partial E}{\partial\ln a_{0}} & = & \frac{\hbar^{2}}{M}\left(\mathcal{I}_{a}^{\left(0\right)}+\frac{\lambda}{\sqrt{2}}\underset{m=\pm1}{\sum}\mathcal{I_{\lambda}^{\left(\mathrm{\mathit{m}}\right)}}\right),\label{eq:2dd1}\\
\frac{\partial E}{\partial a_{1}^{-1}} & = & -\frac{\pi\hbar^{2}}{2M}\underset{m=\pm1}{\sum}\mathcal{I}_{a}^{\left(m\right)},\label{eq:2dd2}\\
\frac{\partial E}{\partial\ln b_{1}} & = & \underset{m=\pm1}{\sum}\left(\mathcal{E}_{m}-\lambda\frac{\hbar^{2}\mathcal{I_{\lambda}^{\left(\mathit{\mathit{m}}\right)}}}{\sqrt{2}M}\right),\label{eq:2dd3}\\
\frac{\partial E}{\partial u} & = & \frac{-\hbar^{2}\lambda}{\sqrt{2}M}\underset{m=\pm1}{\sum}\left[\frac{\mathcal{I_{\lambda}^{\left(\mathit{m}\right)}}}{\sqrt{2}}+\lambda\left(2\mathcal{I}_{a}^{\left(\mathit{m}\right)}+\mathcal{I}_{p}\right)\right],\label{eq:2da6}\\
\frac{\partial E}{\partial v} & = & \frac{\hbar^{2}\lambda}{M}\underset{m=\pm1}{\sum}\mathcal{I}_{\lambda}^{\left(\mathit{\mathrm{\mathit{m}}}\right)}.\label{eq:2da7}
\end{eqnarray}
Obviously, there are additional two new adiabatic energy relations
appear, i.e. Eqs. (\ref{eq:2da6}) and (\ref{eq:2da7}), which originate
from new scattering parameters introduced by SO coupling.

\subsection{Tail of the momentum distribution at large $q$}

In general, the momentum distribution at large $q$ is determined
by the short-range behavior of the many-body wave function when the
fermions $i$ and $j$ are close. Similarly as in the 3D case, we
can formally write the many-body wave function $\Psi_{2D}$ at $\mathbf{r\approx0}$
as the following ansatz
\begin{multline}
\Psi_{2D}\left(\mathbf{X},{\bf r}\right)=\left[\alpha_{0}\ln r+\mathcal{B}_{0}+\mathcal{C}_{0}r^{2}\ln r\right]\Omega_{0}\left(\hat{{\bf r}}\right)\\
+\underset{m}{\sum}\left[\frac{\alpha_{m}}{r}+\mathcal{B}_{m}r\ln r+\mathcal{C}_{m}r\right]\Omega_{m}\left(\hat{{\bf r}}\right)+O\left(r^{2}\right),\label{eq:lm2-1}
\end{multline}
where $\alpha_{j}$, $\mathcal{B}_{j}$ and $\mathcal{C}_{j}$ ($j=0,\pm1$)
are all regular functions of $\mathbf{X}$. Comparing Eqs. (\ref{eq:u2d})
and (\ref{eq:lm2-1}) at small $r$, we find that

\begin{eqnarray}
\mathcal{B}_{0}\left(\mathbf{X}\right) & = & \alpha_{0}\left(\gamma-\ln2a_{0}\right)+\sum_{m=\pm1}\alpha_{m}\lambda u,\\
\mathcal{B}_{m}\left(\mathbf{X}\right) & = & -\frac{\alpha_{m}k^{2}}{2}+\lambda\frac{\alpha_{0}}{\sqrt{2}},\\
\mathcal{C}_{0}\left(\mathbf{X}\right) & = & -\frac{\alpha_{0}k^{2}}{4},\\
\mathcal{C}_{m}\left(\mathbf{X}\right) & = & \alpha_{m}\left(-\frac{\pi}{4a_{1}}+\frac{1-2\gamma}{4}k^{2}\right)\nonumber \\
 &  & +\alpha_{0}\lambda v+\left(\frac{\alpha_{m}k^{2}}{2}-\lambda\frac{\alpha_{0}}{\sqrt{2}}\right)\ln2b_{1}.
\end{eqnarray}
In the follows, we derive the momentum distribution at large $\mathbf{q}$
but still smaller than $\epsilon^{-1}$. With the help of the plane-wave
expansion

\begin{equation}
e^{i\mathbf{q\cdot r}}=\sqrt{2\pi}\stackrel[m=0]{\infty}{\sum}\underset{\sigma=\pm}{\sum}\eta_{m}i^{m}J_{m}\left(qr\right)e^{-i\sigma m\varphi_{q}}\Omega_{m}^{\left(\sigma\right)}\left(\varphi\right),
\end{equation}
where $\eta_{m}=1/2$ for $m=0$, and $\eta_{m}=1$ for $m\geq1$,
and $\varphi_{q}$ is the azimuthal angle of $\mathbf{q}$, we have

\begin{gather}
\int d\mathbf{r}\alpha_{0}\ln r\Omega_{0}\left(\hat{{\bf r}}\right)e^{-i\mathbf{q}\cdot\mathbf{r}}=-\frac{2\pi}{q^{2}}\alpha_{0}\Omega_{0}\left(\hat{\mathbf{q}}\right),\label{eq:2d1}\\
\int d\mathbf{r}\mathcal{B}_{0}\Omega_{0}\left(\hat{{\bf r}}\right)e^{-i\mathbf{q}\cdot\mathbf{r}}=0,\label{eq:2d2}\\
\int d\mathbf{r}\mathcal{C}_{0}r^{2}\ln r\Omega_{0}\left(\hat{{\bf r}}\right)e^{-i\mathbf{q}\cdot\mathbf{r}}=\frac{8\pi}{q^{4}}\mathcal{C}_{0}\Omega_{0}\left(\hat{\mathbf{q}}\right),\label{eq:2d3}\\
\int d\mathbf{r}\frac{\alpha_{m}}{r}\Omega_{m}\left(\hat{{\bf r}}\right)e^{-i\mathbf{q}\cdot\mathbf{r}}=-i\frac{2\pi}{q}\alpha_{m}\Omega_{m}\left(\hat{\mathbf{q}}\right),\label{eq:2d4}\\
\int d\mathbf{r}\mathcal{B}_{m}r\ln r\Omega_{m}\left(\hat{{\bf r}}\right)e^{-i\mathbf{q}\cdot\mathbf{r}}=i\frac{4\pi}{q^{3}}\mathcal{B}_{m}\Omega_{m}\left(\hat{{\bf q}}\right),\label{eq:2d5}\\
\int d\mathbf{r}\mathcal{C}_{m}r\Omega_{m}\left(\hat{{\bf r}}\right)e^{-i\mathbf{q}\cdot\mathbf{r}}=0,\label{eq:2d6}
\end{gather}
where $\hat{\mathbf{q}}$ is the angular part of $\mathbf{q}$. Inserting
Eqs. (\ref{eq:2d1})-(\ref{eq:2d6}) into (\ref{eq:lm1}), we find
that the total momentum distribution $n_{2D}\left(\mathbf{q}\right)$
at large $\mathbf{q}$ takes the form of

\begin{multline}
n_{2D}\left(\mathbf{q}\right)\approx\mathcal{N}\int d\mathbf{X}\underset{m,m^{\prime}}{\sum}\alpha_{m}\alpha_{m^{\prime}}^{*}\Omega_{m}\left(\hat{\mathbf{q}}\right)\Omega_{m^{\prime}}^{*}\left(\hat{\mathbf{q}}\right)\frac{8\pi^{2}}{q^{2}}\\
+i\underset{m}{\sum}\left[\alpha_{0}^{*}\alpha_{m}\Omega_{0}^{*}\left(\hat{\mathbf{q}}\right)\Omega_{m}\left(\hat{\mathbf{q}}\right)-\alpha_{0}\alpha_{m}^{*}\Omega_{0}\left(\hat{\mathbf{q}}\right)\Omega_{m}^{*}\left(\hat{\mathbf{q}}\right)\right]\frac{8\pi^{2}}{q^{3}}\\
+\left\{ \alpha_{0}\alpha_{0}^{*}\Omega_{0}\left(\hat{\mathbf{q}}\right)\Omega_{0}^{*}\left(\hat{\mathbf{q}}\right)+\underset{m,m^{\prime}}{\sum}\left[-\sqrt{2}\lambda\left(\alpha_{0}\alpha_{m}^{*}\Omega_{m^{\prime}}\left(\hat{\mathbf{q}}\right)\Omega_{m}^{*}\left(\hat{\mathbf{q}}\right)\right.\right.\right.\\
\left.\left.\left.+\alpha_{0}^{*}\alpha_{m}\Omega_{m}\left(\hat{\mathbf{q}}\right)\Omega_{m^{\prime}}^{*}\left(\hat{\mathbf{q}}\right)\right)+2k^{2}\alpha_{m}\alpha_{m^{\prime}}^{*}\Omega_{m}\left(\hat{\mathbf{q}}\right)\Omega_{m^{\prime}}^{*}\left(\hat{\mathbf{q}}\right)\right]\right\} \\
\times\frac{8\pi^{2}}{q^{4}}+O\left(q^{-5}\right)\label{eq:n2}
\end{multline}
and the summations are over $m,m^{\prime}=\pm1$. If we are only interested
in the dependence of $n_{2D}\left(\mathbf{q}\right)$ on the amplitude
of $\mathbf{q}$, the expression can further be simplified by integrating
$n_{2D}\left(\mathbf{q}\right)$ over the direction of $\mathbf{q}$,
and all the odd-order terms of $q^{-1}$ vanish. Finally, we arrive
at

\begin{multline}
n_{2D}\left(q\right)=\frac{\underset{m=\pm1}{\sum}C_{a}^{\left(m\right)}}{q^{2}}\\
+\left[C_{a}^{\left(0\right)}+\underset{m=\pm1}{\sum}\left(C_{b}^{\left(m\right)}-\lambda\mathcal{P}_{\lambda}^{\left(m\right)}\right)\right]\frac{1}{q^{4}}+O\left(q^{-6}\right),\label{eq:n2d}
\end{multline}
where the contacts are defined as

\begin{equation}
C_{a}^{\left(j\right)}=8\pi^{2}\mathcal{I}_{a}^{\left(j\right)}\label{eq:2c1}
\end{equation}
for $j=0,\pm1$, and

\begin{eqnarray}
C_{b}^{\left(m\right)} & = & \frac{16\pi^{2}M}{\hbar^{2}}\mathcal{E}_{m},\label{eq:2c2}\\
\mathcal{P}_{\lambda}^{\left(m\right)} & = & 8\sqrt{2}\pi^{2}\mathcal{I}_{\lambda}^{\left(m\right)}\label{eq:2c3}
\end{eqnarray}
for $m=\pm1$. With these definitions in hands, the adiabatic energy
relations (\ref{eq:2dd1})-(\ref{eq:2da7}) can alternatively be written
as

\begin{eqnarray}
\frac{\partial E}{\partial\ln a_{0}} & = & \frac{\hbar^{2}}{8\pi^{2}M}\left(C_{a}^{\left(0\right)}+\frac{\lambda}{2}\underset{m=\pm1}{\sum}\mathcal{P}_{\lambda}^{\left(m\right)}\right),\label{eq:2dc1}\\
\frac{\partial E}{\partial a_{1}^{-1}} & = & -\frac{\hbar^{2}}{16\pi M}\underset{m=\pm1}{\sum}C_{a}^{\left(m\right)},\label{eq:2dc2}\\
\frac{\partial E}{\partial\ln b_{1}} & = & \frac{\hbar^{2}}{16\pi^{2}M}\underset{m=\pm1}{\sum}\left(C_{b}^{\left(m\right)}-\lambda\mathcal{P}_{\lambda}^{\left(m\right)}\right),\label{eq:2dc3}\\
\frac{\partial E}{\partial u} & = & -\frac{\hbar^{2}\lambda}{16\sqrt{2}\pi^{2}M}\underset{m=\pm1}{\sum}\mathcal{P_{\lambda}^{\left(\mathit{m}\right)}},\label{eq:2dc4}\\
\frac{\partial E}{\partial v} & = & \frac{\hbar^{2}\lambda}{8\sqrt{2}\pi^{2}M}\underset{m=\pm1}{\sum}\mathcal{P}_{\lambda}^{\left(m\right)}.\label{eq:2dc5}
\end{eqnarray}
In the absence of SO coupling, Eqs. (\ref{eq:2dc1}), (\ref{eq:2dc2})
and (\ref{eq:2dc3}) simply reduce to the ordinary form of the adiabatic
energy relations for $s$- and $p$-wave interactions \cite{Werner2012GB,Peng2019U},
with respect to the scattering length (or area) as well as effective
range. And for the $s$-wave interaction, there is a difference of
the factor $2\pi$ from the Ref.\cite{Werner2012GB}, which is because
we include the angular part $1/\sqrt{2\pi}$ in the $s$-partial wave
function. In addition, two additional new adiabatic energy relations,
i.e., Eqs. (\ref{eq:2dc4}) and (\ref{eq:2dc5}), and new contacts
$\mathcal{P}_{\lambda}^{\left(m\right)}$ appear, due to SO coupling.

\subsection{The high-frequency tail of the rf spectroscopy}

We may carry out the analogous procedure as that in 3D systems with
3D SO coupling, and the two-body rf transition rate takes the form

\begin{multline}
\Gamma_{2}\left(\omega\right)=\frac{2\pi\gamma_{rf}^{2}}{\hbar}\times\\
\underset{\mathbf{k}_{1}\mathbf{k}_{2}}{\sum}\left(\left|\tilde{\phi}_{\uparrow\downarrow}\right|^{2}+\left|\tilde{\phi}_{\downarrow\uparrow}\right|^{2}+2\left|\tilde{\phi}_{\downarrow\downarrow}\right|^{2}\right)\delta\left(\hbar\omega-\Delta E\right),\label{eq:rf1}
\end{multline}
where

\begin{equation}
\tilde{\phi}_{\sigma_{1}\sigma_{2}}\left(\mathbf{k_{\mathrm{1}}},\mathbf{k}_{2}\right)=\int d{\bf r}_{1}d{\bf r}_{2}\phi_{\sigma_{1}\sigma_{2}}\left({\bf r}_{1},{\bf r}_{2}\right)e^{-i{\bf k}_{1}\cdot{\bf r}_{1}}e^{-i{\bf k}_{2}\cdot{\bf r}_{2}}.\label{eq:2rf2}
\end{equation}
If we are only interested in the high-frequency tail of the transition
rate, we can use the asymptotic behavior of the two-body wave function
for a 2D system with 2D SO coupling, i.e. Eq. (\ref{eq:u2d}). Combining
with Eqs.(\ref{eq:rf1}) and (\ref{eq:2rf2}), we obtain the two-body
rf transition rate $\Gamma_{2}\left(\omega\right)$ as 

\begin{eqnarray}
\Gamma_{2}\left(\omega\right) & = & \frac{M\gamma_{rf}^{2}}{4\pi\hbar^{3}}\left[\frac{c_{a}^{\left(1\right)}}{M\omega/\hbar}\right.\nonumber \\
 &  & \qquad\quad\left.+\frac{c_{a}^{\left(0\right)}/2+c_{b}^{\left(1\right)}/2-\lambda p_{\lambda}^{\left(1\right)}}{\left(M\omega/\hbar\right)^{2}}\right],\label{eq:tr5}
\end{eqnarray}
where $c_{a}^{\left(0\right)}$, $c_{a}^{\left(1\right)}$, $c_{b}^{\left(1\right)}$
and $p_{\lambda}^{\left(1\right)}$ are contacts for a two-body system
with $\mathcal{N=\mathrm{1}}$ in the definitions (\ref{eq:2c1})-(\ref{eq:2c3}). 

For many-body systems, all $\mathcal{N}=N\left(N-1\right)/2$ pairs
contribute to the transition rate. Similarly, we can redefining the
constant $\mathcal{N}$ into the contacts, and then obtain

\begin{eqnarray}
\Gamma_{N}\left(\omega\right) & = & \frac{M\gamma_{rf}^{2}}{4\pi\hbar^{3}}\left[\frac{C_{a}^{\left(1\right)}}{M\omega/\hbar}\right.\nonumber \\
 &  & \qquad\quad\left.+\frac{C_{a}^{\left(0\right)}/2+C_{b}^{\left(1\right)}/2-\lambda\mathcal{P}_{\lambda}^{\left(1\right)}}{\left(M\omega/\hbar\right)^{2}}\right],\label{eq:tr5-1}
\end{eqnarray}
where $C_{a}^{\left(0\right)}$, $C_{a}^{\left(1\right)}$, $C_{b}^{\left(1\right)}$
and $\mathcal{P}_{\lambda}^{\left(1\right)}$are corresponding contacts
for many-body systems. In the absence of SO coupling, Eq. (\ref{eq:tr5-1})
simply reduces to the ordinary results for $s$- and $p$-wave interactions,
respectively \cite{Barth2014P,Peng2019U}.

\subsection{Pair correlation function at short distances}

Let us then discuss the short-distance behavior of the pair correlation
function for a 2D Fermi gases with 2D SO coupling. Inserting the asymptotic
form of the many-body wave function at short distance, i.e. Eq. (\ref{eq:u2d})
into the Eq. (\ref{eq:pc2}), we easily obtain spatially integrated
pair correlation function $G_{2}\left(\mathbf{r}\right)$. If we are
only interested in the dependence of $G_{2}\left(\mathbf{r}\right)$
on the amplitude of $r=\left|\mathbf{r}\right|$, we may integrate
over the direction of $\mathbf{r}$, and obtain

\begin{multline}
G_{2}\left(r\right)\approx\frac{1}{4\pi^{2}}\left[\frac{\underset{m=\pm1}{\sum}C_{a}^{\left(m\right)}}{r^{2}}+C_{a}^{\left(0\right)}\left(\ln\frac{r}{2a_{0}}\right)^{2}\right.\\
+\left(2\gamma C_{a}^{\left(0\right)}+\frac{\lambda u}{\sqrt{2}}\underset{m=\pm1}{\sum}\mathcal{P}_{\lambda}^{\left(m\right)}\right)\ln\frac{r}{2a_{0}}\\
\left.+\underset{m=\pm1}{\sum}\frac{1}{2}\left(-C_{b}^{\left(m\right)}+\lambda\mathcal{P}_{\lambda}^{\left(m\right)}\right)\ln\frac{r}{2b_{1}}\right].\label{eq:mg2}
\end{multline}
In the absence of SO coupling, Eq. (\ref{eq:mg2}) simply reduces
to the ordinary results for $s$- and $p$-wave interactions, respectively
\cite{Werner2012GB,Peng2019U}.

\subsection{Grand canonical potential and pressure relation}

Similarly, according to the dimension analysis, we easily obtain

\begin{eqnarray}
-\mathcal{J}-\frac{a_{0}}{2}\frac{\partial E}{\partial a_{0}}-a_{1}\frac{\partial E}{\partial a_{1}}-\frac{b_{1}}{2}\frac{\partial E}{\partial b_{1}} & = & E.\label{eq:2g1}
\end{eqnarray}
Further by using adiabatic energy relations, Eq.(\ref{eq:2g1}) becomes

\begin{multline}
\mathcal{J}=-E-\frac{\hbar^{2}}{16\pi^{2}M}\left(C_{a}^{\left(0\right)}+\frac{\lambda}{2}\underset{m=\pm1}{\sum}\mathcal{P}_{\lambda}^{\left(m\right)}\right)\\
-\frac{\hbar^{2}}{16\pi M}\underset{m=\pm1}{\sum}\left[\frac{C_{a}^{\left(m\right)}}{a_{1}}+\frac{1}{2\pi}\left(C_{b}^{\left(m\right)}-\lambda\mathcal{P}_{\lambda}^{\left(m\right)}\right)\right].\label{eq:2g2}
\end{multline}
The pressure relation can be obtained by dividing both sides of Eq.(\ref{eq:2g2})
by $-V$, which respectively reduces to the results in the absence
of SO coupling 

\begin{equation}
P=\frac{E}{V}+\frac{\hbar^{2}C_{a}^{\left(0\right)}}{16\pi^{2}MV}\label{eq:g31}
\end{equation}
for $s$-wave interactions, which is consistent with the result of
Ref.\cite{Valiente2011U}, and

\begin{equation}
P=\frac{E}{V}+\underset{m=\pm1}{\sum}\frac{\hbar^{2}}{16\pi MV}\left(\frac{C_{a}^{\left(m\right)}}{a_{1}}+\frac{C_{b}^{\left(m\right)}}{2\pi}\right)\label{eq:g32}
\end{equation}
for $p$-wave interactions, which is consistent with the result of
Ref.\cite{Peng2019U}.

\section{Conclusions}

In conclusion, we systematically study a set of universal relations
for spin-orbit-coupled Fermi gases in three or two dimension, respectively.
The universal short-range forms of two-body wave functions are analytically
derived, by using a perturbation method, in the sub-Hilbert space
of zero center-of-mass momentum and zero total angular momentum of
pairs. The obtained short-range behaviors of two-body wave functions
do not depend on the short-range details of interatomic potentials.
We find that two new microscopic scattering parameters appear because
of spin-orbit coupling, and then new contacts need to be introduced
in both three- and two-dimensional systems. However, due to different
short-range behaviors of two-body wave functions for three- and two-dimensional
systems, the specific forms of universal relations are distinct in
different dimensions. As we anticipate, the universal relations for
spin-orbit-coupled systems, such as the adiabatic energy relations,
the large-momentum distributions, the high-frequency behavior of the
radio-frequency responses, short-range behaviors of the pair correlation
functions, grand canonical potentials, and pressure relations, are
fully captured by the contacts defined. In general, more partial-wave
scatterings should be taken into account for nonzero center-of-mass
momentum and nonzero total angular momentum of pairs. Consequently,
we may expect more contacts to appear. Our results may shed some light
for understanding the profound properties of the few- and many-body
spin-orbit-coupled quantum gases.

\section*{Acknowledgments}

This work has been supported by the NKRDP (National Key Research and
Development Program) under Grant No. 2016YFA0301503, NSFC (Grant No.11674358,
11434015, 11474315) and CAS under Grant No. YJKYYQ20170025.

\section*{Appendix A: normalization of the wave function for 3d systems with
3D SO coupling}

In this section of Appendix A, we are going to derive $\int_{\mathcal{D}_{\epsilon}}\stackrel[i=1]{N}{\prod}d\mathbf{r_{\mathit{i}}}\left|\Psi\right|^{2}$
for 3D many-body systems with 3D SO coupling. Let us consider two
many-body wave functions $\Psi^{\prime}$ and $\Psi$, corresponding
to different energies $\hbar^{2}k^{\prime2}/M$ and $\hbar^{2}k^{2}/M$,
respectively. They should be orthogonal, i.e., $\int_{\mathcal{D}_{\epsilon}}\stackrel[i=1]{N}{\prod}d\mathbf{r_{\mathit{i}}}\Psi^{\prime*}\Psi=0$,
and therefore we have

\begin{equation}
\int_{r<\epsilon}\stackrel[i=1]{N}{\prod}d\mathbf{r_{\mathit{i}}}\Psi^{\prime*}\Psi=-\int_{r>\epsilon}\stackrel[i=1]{N}{\prod}d\mathbf{r_{\mathit{i}}}\Psi^{\prime*}\Psi.
\end{equation}
From the Schr\"{o}dinger equation satisfied by $\Psi^{\prime}$ and
$\Psi$ outside the interaction potential, i.e., $r>\epsilon$, we
easily obtain

\begin{multline}
\int_{r>\epsilon}\stackrel[i=1]{N}{\prod}d\mathbf{r_{\mathit{i}}}\Psi^{\prime*}\Psi=\frac{\epsilon^{2}}{k^{2}-k^{\prime2}}\mathcal{N}\int d\mathbf{X}\int_{r=\epsilon}d\mathbf{\hat{r}}\\
\left[\left(\Psi^{\prime*}\frac{\partial}{\partial r}\Psi-\Psi\frac{\partial}{\partial r}\Psi^{\prime*}\right)+\frac{\lambda}{2\pi}\left(\psi_{0}^{\prime*}\psi_{1}-\psi_{1}^{\prime*}\psi_{0}\right)\right].
\end{multline}
In the presence of SO coupling, only $s$- and $p$-wave scatterings
are involved in the subspace $\mathbf{K}=0$ and $\mathbf{J}=0$,
and the wave function at short distance takes the form of Eq. (\ref{eq:f11}).
Using the asymptotic behavior of the wave function, we easily evaluate

\begin{alignat}{1}
 & \int_{r<\epsilon}\stackrel[i=1]{N}{\prod}d\mathbf{r_{\mathit{i}}}\left|\Psi\right|^{2}\nonumber \\
 & =-\underset{k^{\prime}\rightarrow k}{\lim}\frac{1}{2}\left(\int_{r>\epsilon}\stackrel[i=1]{N}{\prod}d\mathbf{r_{\mathit{i}}}\Psi^{\prime*}\Psi+\int_{r>\epsilon}\stackrel[i=1]{N}{\prod}d\mathbf{r_{\mathit{i}}}\Psi^{\prime}\Psi^{*}\right)\nonumber \\
 & =-\mathcal{N}\int d\mathbf{X}\left\{ \frac{\left|\alpha_{1}\right|^{2}}{\epsilon}\left.+\frac{\left|\alpha_{1}\right|^{2}b_{1}}{2}\right\} \right.\nonumber \\
 & =-\left(\frac{1}{\epsilon}+\frac{b_{1}}{2}\right)\mathcal{I}_{a}^{\left(1\right)},
\end{alignat}
which in turn yields

\begin{multline}
\int_{\mathcal{D}_{\epsilon}}\stackrel[i=1]{N}{\prod}d\mathbf{r_{\mathit{i}}}\left|\Psi\right|^{2}=1+\left(\frac{1}{\epsilon}+\frac{b_{1}}{2}\right)\mathcal{I}_{a}^{\left(1\right)}.
\end{multline}

\section*{Appendix B: normalization of the wave function for 2d system with
2d SO coupling}

In this section of Appendix B, we are going to derive $\int_{\mathcal{D}_{\epsilon}}\stackrel[i=1]{N}{\prod}d\mathbf{r_{\mathit{i}}}\left|\Psi\right|^{2}$
for 2D many-body systems with 2D SO coupling. Let us consider two
many-body wave functions $\Psi^{\prime}$ and $\Psi$, corresponding
to different energies $\hbar^{2}k^{\prime2}/M$ and $\hbar^{2}k^{2}/M$,
respectively. They should be orthogonal, i.e., $\int_{\mathcal{D}_{\epsilon}}\stackrel[i=1]{N}{\prod}d\mathbf{r_{\mathit{i}}}\Psi^{\prime*}\Psi=0$,
and therefore we have

\begin{equation}
\int_{r<\epsilon}\stackrel[i=1]{N}{\prod}d\mathbf{r_{\mathit{i}}}\Psi^{\prime*}\Psi=-\int_{r>\epsilon}\stackrel[i=1]{N}{\prod}d\mathbf{r_{\mathit{i}}}\Psi^{\prime*}\Psi.
\end{equation}
From the Schr\"{o}dinger equation satisfied by $\Psi^{\prime}$ and
$\Psi$ outside the interaction potential, i.e., $r>\epsilon$, we
easily obtain

\begin{multline}
\int_{r>\epsilon}\stackrel[i=1]{N}{\prod}d\mathbf{r_{\mathit{i}}}\Psi^{\prime*}\Psi=\frac{\epsilon}{k^{2}-k^{\prime2}}\mathcal{N}\int d\mathbf{X}\int_{r=\epsilon}d\mathbf{\hat{r}}\\
\left[\left(\Psi^{\prime*}\frac{\partial}{\partial r}\Psi-\Psi\frac{\partial}{\partial r}\Psi^{\prime*}\right)+\underset{m=\pm1}{\sum}\frac{\lambda}{\sqrt{2}\pi}\left(\psi_{0}^{\prime*}\psi_{m}-\psi_{m}^{\prime*}\psi_{0}\right)\right].
\end{multline}
In the presence of SO coupling, only $s$- and $p$-wave scatterings
are involved in the subspace $\mathbf{K}=0$ and $\mathbf{J}=0$,
and the wave function at short distance takes the form of Eq. (\ref{eq:u2d}).
Using the asymptotic behavior of the wave function, we easily evaluate

\begin{align}
 & \int_{r<\epsilon}\stackrel[i=1]{N}{\prod}d\mathbf{r_{\mathit{i}}}\left|\Psi\right|^{2}\nonumber \\
 & =-\underset{k^{\prime}\rightarrow k}{\lim}\frac{1}{2}\left(\int_{r>\epsilon}\stackrel[i=1]{N}{\prod}d\mathbf{r_{\mathit{i}}}\Psi^{\prime*}\Psi+\int_{r>\epsilon}\stackrel[i=1]{N}{\prod}d\mathbf{r_{\mathit{i}}}\Psi^{\prime}\Psi^{*}\right)\nonumber \\
 & =\mathcal{N}\int d\mathbf{X}\underset{m=\pm1}{\sum}\left(\ln\frac{\epsilon}{2b_{1}}+\gamma\right)\left|\alpha_{m}\right|^{2}\nonumber \\
 & =\underset{m=\pm1}{\sum}\left(\ln\frac{\epsilon}{2b_{1}}+\gamma\right)\mathcal{I}_{a}^{\left(m\right)},
\end{align}
which in turn yields

\begin{multline}
\int_{\mathcal{D}_{\epsilon}}\stackrel[i=1]{N}{\prod}d\mathbf{r_{\mathit{i}}}\left|\Psi\right|^{2}=1-\underset{m=\pm1}{\sum}\left(\ln\frac{\epsilon}{2b_{1}}+\gamma\right)\mathcal{I}_{a}^{\left(m\right)}.
\end{multline}

\end{document}